\documentclass[aps,prc,twocolumn,superscriptaddress,lineno]{revtex4-2}   %%%%%%%%% two columns

\usepackage[english]{babel}

\usepackage{lineno}

\usepackage{isotope}
\usepackage[final]{todonotes}

\usepackage{amsmath}
\usepackage{amssymb}
\usepackage{multirow}
\newcommand\hashtag{\scalebox{0.8}{\raisebox{0.4ex}{\#}}}
\usepackage{float}
\usepackage{soul} 
\usepackage{graphicx}

\usepackage{lipsum}
\usepackage[export]{adjustbox} 

\usepackage{url}

\usepackage{hyperref}
\hypersetup{
	colorlinks=true,   % Activa los enlaces de color
	linkcolor=blue,    % Color de enlaces internos
	citecolor=blue,    % Color de enlaces de citas
	filecolor=blue,    % Color de enlaces a archivos
	urlcolor=blue      % Color de enlaces URL
}
\usepackage{xpatch}
\usepackage{pdfcomment}  % Para tooltips

\xpatchbibdriver{article}
{\usebibmacro{journal+issuetitle}}
{\printfield{journaltitle}%
	\iffieldundef{doi}{}%
	{\pdfcomment{\printfield{journaltitle}}{DOI: \url{https://doi.org/\thefield{doi}}}}}
{}{}

\makeatletter

\def\p@subsection{}
\def\p@subsubsection{}
\makeatother

\begin{document}

	%\preprint{asd}

	\title{ The beta decay of $\mathbf{T_z=-2}$ \isotope[\bf 64]{\bf Se} and its descendants: the T=2 isobaric multiplet 
	}

	
	
	\author{P. Aguilera}
	\email{Corresponding author: pablo.aguilera@pd.infn.it}
	\affiliation{Instituto de F\'isica Corpuscular, CSIC-Universitat de Val\'encia, E-46071 Val\'encia, Spain}
	\affiliation{Centro de Investigación en F\'isica Nuclear y Espectroscop\'ia de Neutrones CEFNEN, Comisi\'on Chilena de Energ\'ia Nuclear, Nueva Bilbao 12501, Las Condes, Santiago, Chile}
	\affiliation{Istituto Nazionale di Fisica Nucleare, Sezione di Padova, Padova I-35131, Italy}
	%\affiliation{Chilean Nuclear Energy Commission}
	%\affiliation
	
	\author{F. Molina}
	\affiliation{Centro de Investigación en F\'isica Nuclear y Espectroscop\'ia de Neutrones CEFNEN, Comisi\'on Chilena de Energ\'ia Nuclear, Nueva Bilbao 12501, Las Condes, Santiago, Chile}
	\affiliation{Millennium Institute for Subatomic physics at high energy frontier - SAPHIR, Fern\'andez Concha 700, Las Condes, Santiago, Chile}
	\affiliation{Departamento de Ciencias F\'isicas, Universidad Andres Bello, Sazi\'e 2212, 837-0136, Santiago, Chile}
	
	\author{B. Rubio}
	\email{Corresponding author: berta.rubio@ific.uv.es}
	\affiliation{Instituto de F\'isica Corpuscular, CSIC-Universitat de Val\'encia, E-46071 Val\'encia, Spain}
	\author{S.E.A. Orrigo}
	\affiliation{Instituto de F\'isica Corpuscular, CSIC-Universitat de Val\'encia, E-46071 Val\'encia, Spain}
	
	\author{W. Gelletly}
	\affiliation{Instituto de F\'isica Corpuscular, CSIC-Universitat de Val\'encia, E-46071 Val\'encia, Spain}
	\affiliation{Department of Physics, University of Surrey, Guildford GU2 7XH, United Kingdom}
	
	\author{Y. Fujita}
	\affiliation{Research Center for Nuclear Physics, Osaka University, Ibaraki, Osaka 567-0047, Japan}
	\affiliation{Department of Physics, Osaka University, Toyonaka, Osaka 560-0043, Japan}
	
	\author{J. Agramunt}
	\affiliation{Instituto de F\'isica Corpuscular, CSIC-Universitat de Val\'encia, E-46071 Val\'encia, Spain}
	\author{A. Algora}
	\affiliation{Instituto de F\'isica Corpuscular, CSIC-Universitat de Val\'encia, E-46071 Val\'encia, Spain}
	\affiliation{HUN-REN Institute for Nuclear Research (ATOMKI), H-4001 Debrecen, Hungary}
	\author{V. Guadilla}
	\affiliation{Instituto de F\'isica Corpuscular, CSIC-Universitat de Val\'encia, E-46071 Val\'encia, Spain}
	\affiliation{Faculty of Physics, University of Warsaw, 02-093, Warsaw, Poland}
	\author{A. Montaner-Piz\'a}
	\author{A.I. Morales}
	\affiliation{Instituto de F\'isica Corpuscular, CSIC-Universitat de Val\'encia, E-46071 Val\'encia, Spain}

	\author{H.F. Arellano}
	\affiliation{Universidad de Chile, Facultad de Ciencias F\'isicas y Matem\'aticas, Av. Blanco
		Encalada 2008, Santiago, Chile.}
	
	\author{P. Ascher}
	\author{B. Blank}
	\author{M. Gerbaux}
	\author{J. Giovinazzo}
	\author{T. Goigoux}
	\author{S. Gr\'evy}
	\author{T. Kurtukian Nieto}
	\author{C. Magron}
	\affiliation{Centre d'Etudes Nucl\'eaires de Bordeaux-Gradignan, UMR 5797 Universit\'e de Bordeaux CNRS/IN2P3, 19 Chemin du Solarium, CS10120, 33175 Gradignan Cedex, France}
	
	\author{J. Chiba}
	\author{D. Nishimura}
	\author{S. Yagi}
	\affiliation{Department of Natural Sciences, Tokyo City University, Setagaya-ku, Tokyo 158-8557, Japan}
	
	\author{H. Oikawa}
	\author{Y. Takei}
	\affiliation{Department of Physics, Tokyo University of Science, Noda, Chiba 278-8510, Japan}
	
	\author{D.S. Ahn}
	\author{P. Doornenbal}
	\author{N. Fukuda}
	\author{N. Inabe}
	\author{T. Kubo}
	\author{S. Kubono}
	\author{S. Nishimura}
	\author{Y. Shimizu}
	% \author{C. Sidong}   % Asked to be removed
	\author{T. Sumikama}
	\author{H. Suzuki}
	\author{H. Takeda}
	\author{J. Wu}
	\affiliation{RIKEN Nishina Center, 2-1 Hirosawa, Wako, Saitama 351-0198, Japan}
	
	\author{V.H. Phong}
	\affiliation{Faculty of Physics, VNU Hanoi University of Science, 334 Nguyen Trai, Thanh Xuan, Hanoi, Vietnam}
	\affiliation{RIKEN Nishina Center, 2-1 Hirosawa, Wako, Saitama 351-0198, Japan}
	
	\author{G.G. Kiss}
	\affiliation{RIKEN Nishina Center, 2-1 Hirosawa, Wako, Saitama 351-0198, Japan}
	\affiliation{HUN-REN Institute for Nuclear Research, Bem t\'er 18/c, Debrecen, 4026, Hungary}
	
	\author{P.-A. S\"oderstr\"om}
	\affiliation{RIKEN Nishina Center, 2-1 Hirosawa, Wako, Saitama 351-0198, Japan}
	\affiliation{Extreme Light Infrastructure-Nuclear Physics (ELI-NP)/Horia Hulubei National Institute for Physics and Nuclear Engineering (IFIN-HH), Str. Reactorulu, M\u{a}gurele 077125, Romania}

	\author{M. Tanaka}
	\affiliation{Department of Physics, Osaka University, Toyonaka, Osaka 560-0043, Japan}
	
	\author{F. Diel}
	\affiliation{Institute of Nuclear Physics, University of Cologne, D-50937 Cologne, Germany}
	
	\author{D. Lubos}
	\affiliation{Physik Department E12, Technische Universität M\"unchen, D-85748 Garching, Germany}
	
	\author{S. Lenzi}
	\affiliation{Dipartimento di Fisica dell' Universit\`a degli Studi di Padova, Padova I-35131, Italy}
	\affiliation{Istituto Nazionale di Fisica Nucleare, Sezione di Padova, Padova I-35131, Italy}
	
	\author{G. de Angelis}
	\author{D. Napoli}
	\affiliation{Istituto Nazionale di Fisica Nucleare, Laboratori Nazionali di Legnaro, Legnaro I-35020, Italy}
	
	\author{C. Borcea}
	\affiliation{National Institute for Physics and Nuclear Engineering IFIN-HH, P.O. Box MG-6, Bucharest-Magurele, Romania}
	
	\author{A. Boso}
	\affiliation{Dipartimento di Fisica dell' Universit\`a degli Studi di Padova, Padova I-35131, Italy}
	\affiliation{Istituto Nazionale di Fisica Nucleare, Sezione di Padova, Padova I-35131, Italy}
	
	\author{R.B. Cakirli}
	\affiliation{Department of Physics, Istanbul University, Istanbul 34134, Turkey}
	\affiliation{Max Planck Institute for Nuclear Physics, Saupfercheckweg 1, 69117 Heidelberg, Germany.}
	\affiliation{ExtreMe Matter Institute EMMI, GSI Helmholtzzentrum für Schwerionenforschung GmbH, Planckstraße 1, 64291 Darmstadt, Germany.}
	
	\author{E. Ganioglu}
	\affiliation{Department of Physics, Istanbul University, Istanbul 34134, Turkey}
	
	\author{G. de France}
	\affiliation{Grand Acc\'el\'erateur National d'Ions Lourds, B.P. 55027, F-14076 Caen Cedex 05, France}
	
	\author{S. Go}
	\affiliation{Department of Physics, Tennessee University}

	
	
	
	\date{\today}

	\begin{abstract}
		In this paper we present our results on the decay of \isotope[64]{Se}. It is the heaviest $T_z\!=\!-2$ nucleus that both beta decays and has a stable mirror partner ($T_z\!=\!+2$), thus allowing comparison with charge exchange reaction studies. 
		The beta decays of  \isotope[64]{Se} and its descendants were studied at the RIKEN Nishina Center (Tokyo, Japan) following their production in the fragmentation of \isotope[78]{Kr} on a beryllium target. 
		Beta-delayed $\gamma$-ray and particle radiation was identified for each of the nuclei in the decay chain allowing us to obtain  decay schemes for \isotope[64]{Se}, \isotope[64]{As}, and \isotope[63]{Ge}. 
		Thus new excited states could be found for the descendant nuclei, including the interesting case of the $N\!=
		\!Z$ nucleus \isotope[64]{Ge}. 
		Furthermore we observed for the first time the beta-delayed proton emission of \isotope[64]{Se}  and \isotope[64]{As}. 
		Based on these results we obtained proton branching ratios of $ 48.0(9)\% $ in \isotope[64]{Se} decay and $ 4.4(1)\% $ in \isotope[64]{As} decay. 
		We obtained a half-life value of $ 22.5(6)\mathrm{~ms} $ for \isotope[64]{Se} decay and half-lives slightly more precise than those in the literature for each nucleus involved in the decay chain. 
		Using our results on the excited levels of \isotope[64]{As} and the  mass excess in the literature for \isotope[63]{Ge} we obtained $ -39588(50)\mathrm{~keV} $ for the mass excess of \isotope[64]{As}.
		Then based on the IMME we obtained the mass excess of $ -27429(88)\mathrm{~keV} $ for \isotope[64]{Se} by extrapolation. 
		The mirror process of \isotope[64]{Se} beta decay, the charge exchange reaction ${}^{64}\mathrm{Zn}(^{3}\mathrm{He},t)^{64}\mathrm{Ga}$, has already been measured allowing us to study the mirror symmetry through the comparison of the weak force (beta decay) and strong force (charge exchange reaction). An interpretation of the decay schemes based on the idea of the Anti Analogue State is proposed.
	\end{abstract}

	%\keywords{}
	
	\maketitle

	\section{Introduction}
	\label{sec:intro}

	Beta-decay studies far from the line of nuclear stability are a very powerful tool to determine the details of nuclear structure in very unstable nuclei. 
	This is because the overlaps  between the  parent state wave function and the wave functions of states populated in the daughter nucleus are often large.  
	This is particularly true in the case of the Isobaric Analogue State (IAS), populated in Fermi (F) decay, where parent and daughter states are identical in structure except for the third component of the isospin, but it is also true, for daughter states strongly populated in Gamow-Teller (GT) transitions. 
	Moreover, in nuclei far from the line of stability the energy window accessible in $ \beta $ decay is large providing information on nuclear states at high energy. In the study of the decay of proton-rich nuclei there is another interesting aspect, namely, that in some cases there is the possibility of comparing mirror nuclei where the nucleus populated in beta decay is far from stability and its  mirror can be populated in Charge Exchange reactions, the equivalent process to beta decay under certain experimental conditions. Hence we can check the validity of isospin symmetry when one of the mirror nuclei is loosely bound and the other is not. 
	Moreover, we can compare, not only mirror nuclei but mirror processes. This idea has become particularly fruitful in recent years because of advances, both in accessing very exotic $ \beta^+ $ decaying species and also because of the high resolution achieved in the $(p,n)$-type charge exchange reactions that are possible at the Osaka RCNP Grand Raiden \cite{Fujiwara1999} spectrometer facility using the $ (\isotope[3]{He},t) $ reaction. 
	In addition, studies of the $ \beta^+ $ decay of  very exotic nuclei with short half lives have been carried out at several fragmentation facilities with sophisticated experimental techniques to detect beta-delayed proton emission and $\gamma$-rays.

	%%%%%%%%%%%%%Fig.1 Nuclide table%%%%%%%%%%%%%%%%%%%%
	\begin{figure*}[!t]
		\includegraphics[width=17.2cm]{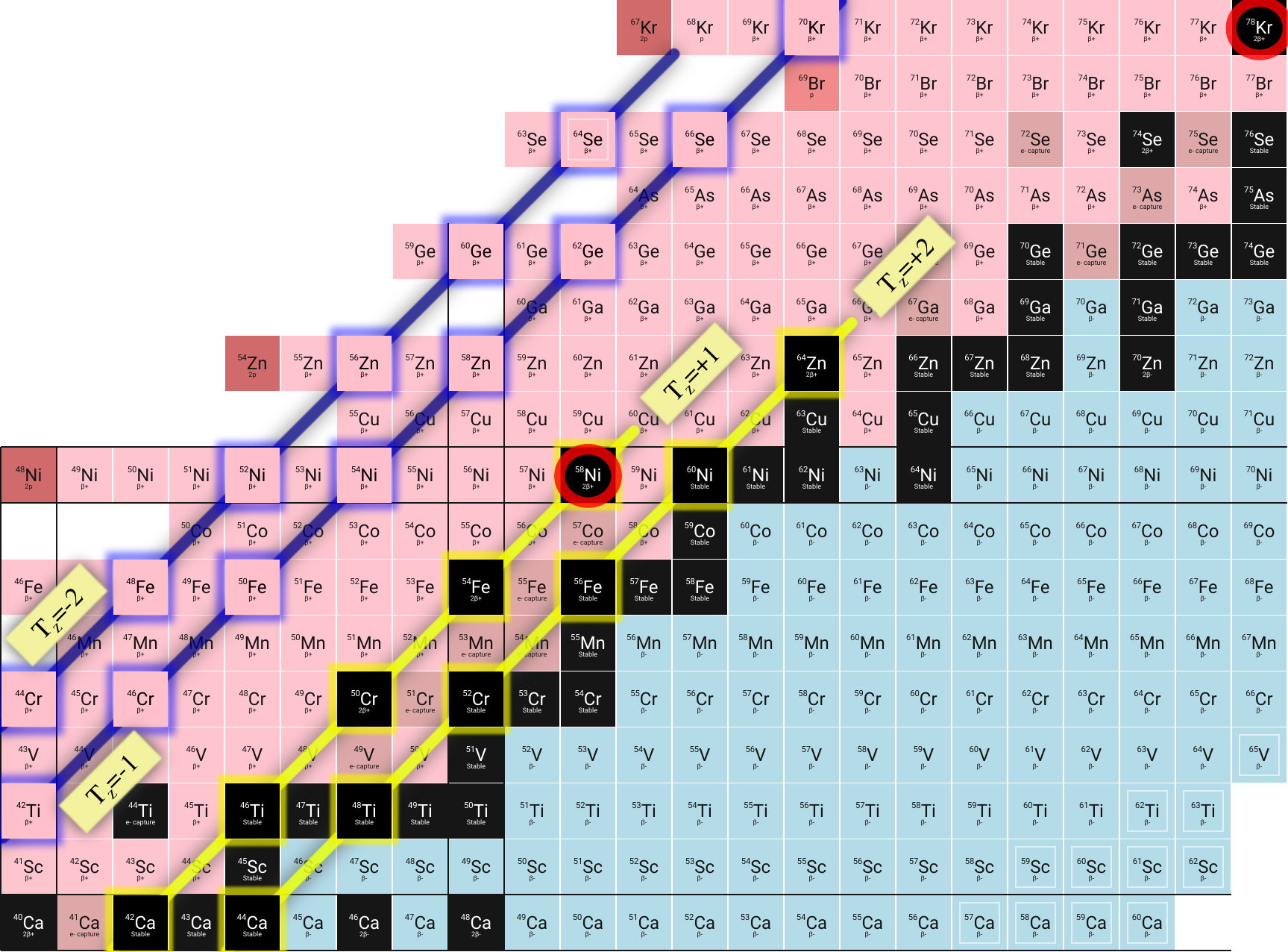}%
		\caption{Section of the chart of the nuclides. Blue circles indicate the even-even $T_{z}\!=\!-2$ and -1 nuclei where the beta decay has been studied in previous campaigns at GSI and GANIL (see text) using the fragmentation of \isotope[58]{Ni} beams, marked in purple, as well as in the present campaign at RIKEN using \isotope[78]{Kr} beams, also in purple. Yellow circles indicate the corresponding  stable mirror nuclei where CE reactions can be carried out. The main focus of the present article is the study of the beta decay of the \isotope[64]{Se} isotope and its descendants. \label{fig:fig_1_nuclides}}
	\end{figure*}

	We have performed studies of this kind on a range of $T_z=\pm $1 and $T_z=\pm $2 even-even nuclei at GSI, GANIL and RIKEN. The results can be seen in the following publications \cite{Orr14, Orrigo2016a,  Mol15, kucuk2017, orrigo2021, Vit22,aguilera_jorquera_study_2020}, and a summary of all the cases studied is shown in Fig. \ref{fig:fig_1_nuclides}. The beta decays of nuclei up to the Ni isotopes, could be produced using the fragmentation of \isotope[58]{Ni} ions on a Be target at GANIL or on a natural Ni target at GSI.
	These  experiments completed the study of beta decay up to the point where the protons have filled the 
	f$_{7/2}$ shell. At GANIL it was possible to go a little beyond that to produce the  \isotope[56]{Zn} and 
	\isotope[58]{Zn} nuclei using either \isotope[58]{Ni} or 
	\isotope[64]{Zn} \cite{Orr14, kucuk2017}  beams. However in order to study
	heavier cases, a primary beam of  \isotope[78]{Kr} was necessary, which is both further away from the nuclei of interest and further away
	from the N$=$Z line in comparison with \isotope[58]{Ni}. As a consequence, they could only be produced with a much smaller cross section. In order to compensate for the resulting reduction in the production rate we had to go to the facility with the highest possible beam intensity, which was the accelerator complex at the RIKEN Nishina Centre in Japan at the time of this experiment.

	The results of studies of \isotope[60]{Ge}, \isotope[62]{Ge} and \isotope[70]{Kr} decay have already been published \cite{orrigo2021, Vit22}. The $ T_z=-2 $ nucleus \isotope[68]{Kr} decays mainly by $\beta$ delayed protons  \cite{giovinazzo2020}, but at present only the half-life and the proton branching is known. In the present  paper we present our results on \isotope[64]{Se}, the heaviest $ T_z=-2 $ nucleus that both undergoes beta decays and fulfils the condition that the comparison with the mirror nucleus is still possible because there is a stable target to allow the study of charge exchange. 
	
	The $ T_z\!=\!-2 $ \isotope[64]{Se} nucleus was first observed in 2005 \cite{Stolz2005} at MSU with a total of four nuclei observed.
	No knowledge of the decay scheme was available prior to the present study. In our experiment we could select the decay of this nucleus and the decay of the daughter nuclei, \isotope[64]{As} and \isotope[63]{Ge}, which was very useful in helping to identify the $\beta$-delayed radiation. 
	More detailed data can be found  in the PhD thesis \cite{aguilera_jorquera_study_2020} and preliminary results were presented in the conference proceedings \cite{rubio_beta_2019}.
	
	This article is organised as follows. In Section \ref{sec:experiment} the experiment performed at the RIKEN Nishina Center is discussed together with a summary of the analysis. Then, in Section \ref{sec:masses} we discuss the ground state masses of \isotope[64]{As} and \isotope[64]{Se}.  In Section \ref{sec:discussions} we present the decay schemes of \isotope[64]{Se}, \isotope[64]{As} and \isotope[63]{Ge}. In Section \ref{sec:multiplet discussion} we discuss the physics of the full $T\!=\!2$ multiplet and in Section \ref{sec:analysis_63ge} the decay of \isotope[63]{Ge} to its mirror \isotope[63]{Ga}. Finally in Section \ref{sec:conclusions} we summarise the conclusions.

	%%%%%%%%%%%%% Section Experiment %%%%%%%%%%%%%%%%%
	%****************************************************
	\section{The experiment and summary of the analysis}
	\label{sec:experiment}
	%****************************************************
	\subsection{Experiment}
	%****************************************************

\begin{figure}
		\includegraphics[width=8.6cm]{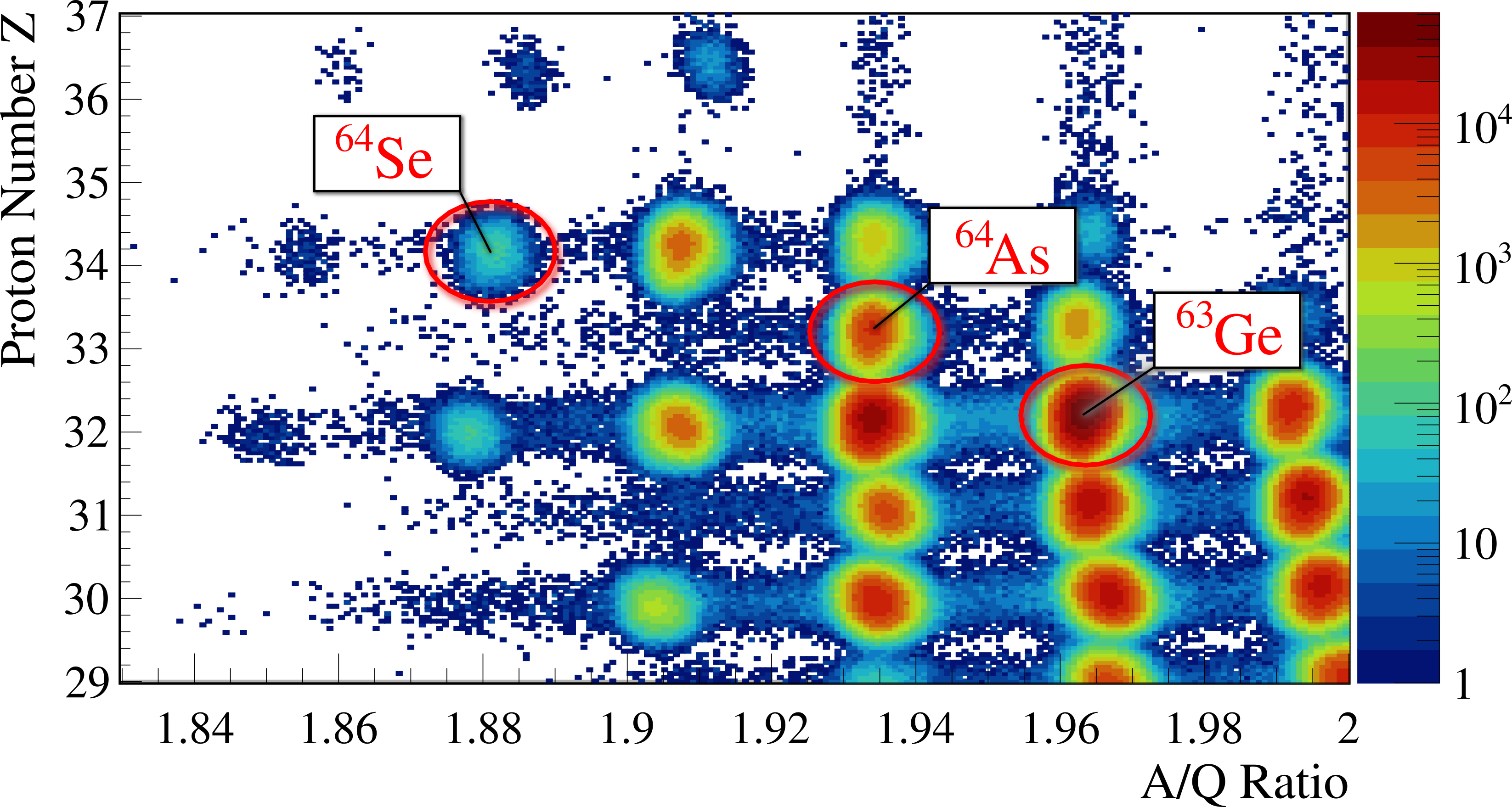}
		\caption{ The proton number $ Z $ versus the mass over charge ratio $ A/Q $ for the fragments produced in the fragmentation of 345 MeV/u \isotope[78]{Kr} ions on a natural Be target. The nuclei studied in this work are labelled. 
			\label{fig:pid_3001}
		}
	\end{figure}

	The experiment was performed at the Radioactive Isotope Beam  Factory (RIBF) at the RIKEN Nishina Centre using the fragmentation of \isotope[78]{Kr} ions of $345$ MeV/u on a $^\mathrm{nat}\mathrm{Be}$ target with an unprecedented beam intensity of up to $300$ pnA. This beam was produced by the RILAC2-RRC-fRC-IRC-SRC combination of accelerators \cite{Okuno2020} at RIBF \cite{injectors_riken_sakamoto,FUK2013,KUB2012}. 
	The fragments produced were separated and identified with the BigRIPS fragment separator  \cite{KUB2012} using the mass-over-charge ratio $ A/Q $  and the proton number $ Z $, obtained using the TOF-$B\rho$-$\Delta E$ method \cite{FUK2013}. 
	The separation in terms of $ Z $ and $ A/Q $ is shown in Fig. \ref{fig:pid_3001} for the present measurement, where nuclei studied in this work are labelled.

	%%%%%%%%%%%% Fig 2 ID plot %%%%%%%%%%%%

	The fragments were conducted to the Zero degree spectrometer placed at the last focal plane of BigRIPS (F11)  to be implanted in the WAS3ABi active stopper \cite{Nishimura2012}. 
	WAS3ABi consisted of three layers of $ 1\mathrm{~mm}
	$ thick Double Sided Silicon Strip Detectors (DSSSD) with an active area of $60\!\times\!40\mathrm{~mm^2}$, segmented into 60 vertical and 40 horizontal strips, defining pixels of dimensions $1\mathrm{~mm}\!\times\!1\mathrm{~mm}\!\times\!1\mathrm{~mm}$. 
	The WAS3ABi active stopper was surrounded by the EUROBALL-RIKEN Cluster Array (EURICA) \cite{SOD2013}, an array of twelve CLUSTER-detectors of EUROBALL type \cite{eberth_development_1997}, each of which consisted of seven tapered hexagonal HPGe crystals. 
	Each cluster was placed at an average distance of $ 22\mathrm{~cm} $ from the centre of WAS3ABi. 
	The trigger signal was a logic OR between the X-side of the DSSSDs and the plastic scintillator at the  F11 focal plane. The 
	$\gamma$ CLUSTER signals were recorded in prompt coincidence
	with the DSSSD signal.

	The fragments were identified off-line using the Z and A/Q ratio. 
	To determine the implantation position at the DSSSD the following criteria were applied: (a) The plastic scintillator at the last focal plane F11 indicates when an ion arrives at the Zero Degree spectrometer, (b) the last DSSSD with an energy overflow indicates in which of the three DSSSD the implantation has occurred and (c) a signal in the plastic detector  placed after WAS3ABi indicates the ions that were not implanted in any DSSSD and acts as a veto. The implantation position in the DSSSD in  X and Y  was determined from the pairs of strips with the fastest time signal on each side of the DSSSD. A specific implanted nucleus is selected by imposing conditions on the Z and mass over charge ratio (A/Q). According to this criterion about $9.0\times10^4$ implanted \isotope[64]{Se} ions were detected in the central DSSSD of WAS3ABi.
	
	The decay events involving positrons and protons were selected as those giving a signal in at least one DSSSD but not in the F11 plastic detector of the focal plane. 
	
	Since the range of protons with energies of a few MeV in silicon is about $ 100\mathrm{~\mu m} $ it was possible to collect all of the energy in one single strip. 
	On the other hand, positrons deposit only a small part of their energy in one strip,  thus making the reconstruction of the positron  emission position difficult. Consequently several pixels,  where a combination of X and Y strips fired, were considered to be involved in the decay event.
	
	To obtain the beta, $\gamma$ and proton spectra for the decay of each nucleus, time correlations between an implantation event and all the decays within a maximum time window of seven times the half-life of the mother nucleus were constructed. 
	For a given implant in one pixel, an accepted decay has to be in the same pixel or in any of the adjacent 8 pixels. This drastically reduces the random correlation background. The full energy ranges for the charged particles in the X and Y strips were 4 MeV and 10 MeV.   
	The energy of the decay signal was chosen to be the maximum energy of all signals registered in the  Y strips of this event. See \cite{morales_simultaneous_2017} for more details.

	%%%%%%%% Subsectio Decay Scheme %%%%%%%%%%%%%%%
	%****************************************************
	\subsection{Decay schemes}
	\label{sec:decay_schemes}
	%****************************************************
	
\begin{figure}
		\includegraphics[width=6.5cm]{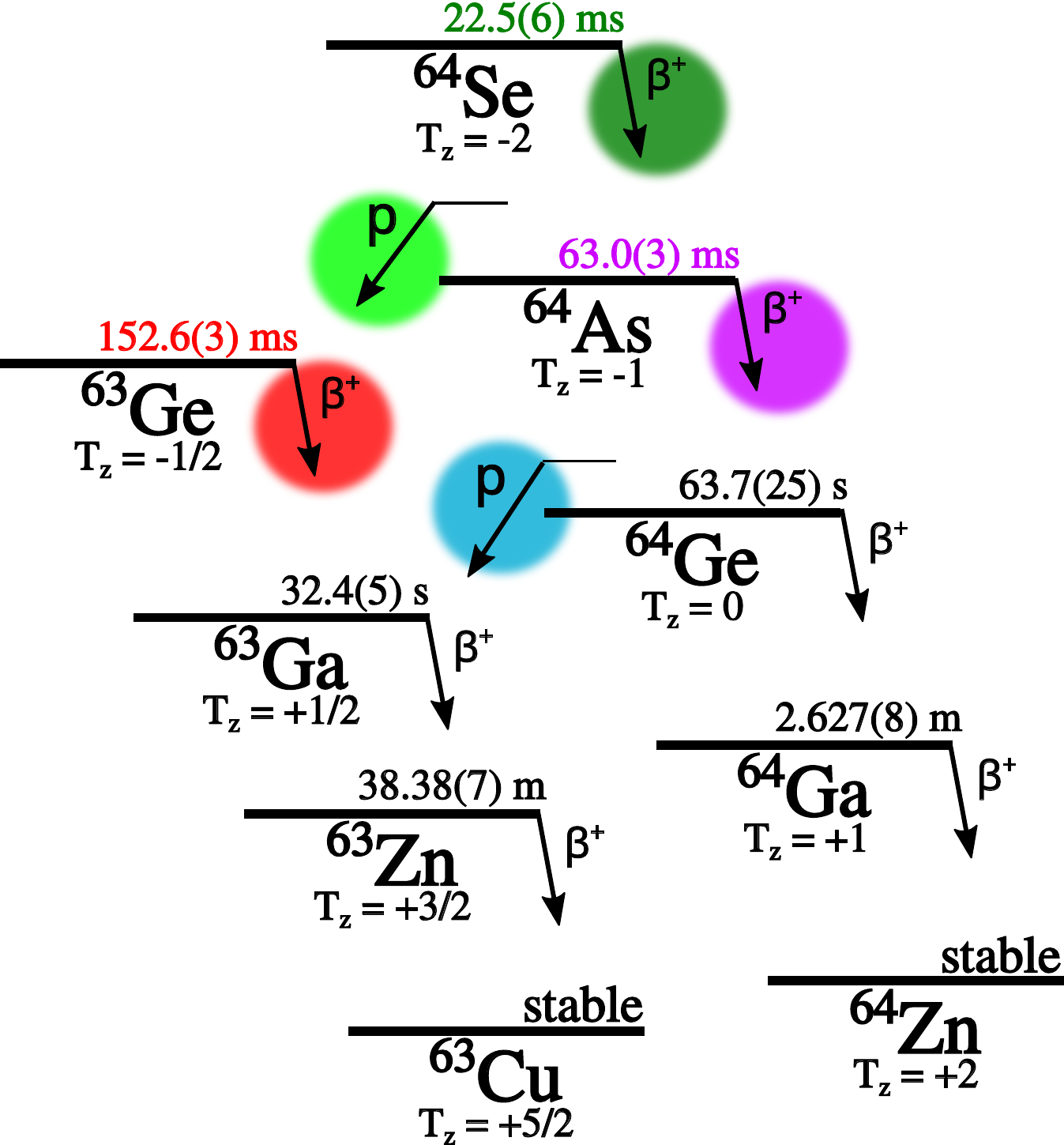}%
		\caption{ The radioactive decay chain starting with  \isotope[64]{Se}. 
			The decays with half-lives shorter than $200\mathrm{~ms}$ are highlighted with a circle. Known half-lives are shown in black, while the half-lives determined in this work are shown in color.\label{fig:decay_scheme_colors}
		}
	\end{figure}

	As mentioned above, no experimental information on the decay properties of  \isotope[64]{Se} existed prior to this work. Accordingly, it was necessary to understand all of the decays correlated with the implantation of \isotope[64]{Se} ions to determine the primary radiation due to \isotope[64]{Se} decay. The decay chain associated with \isotope[64]{Se} is shown in Fig. \ref{fig:decay_scheme_colors} where decays shorter than $200 \mathrm{~ms}$ are marked in colour. 
	A strong beta-delayed proton branching is expected in the decay of \isotope[64]{Se} since the proton separation energy in the daughter nucleus  \isotope[64]{As} is expected to be of the order of $-100\mathrm{~keV}$ according to the most recent Atomic Mass Evaluation (AME2020) tables \cite{wang_ame_2021}.
	In consequence, three beta decays are important here, namely the nucleus of interest, \isotope[64]{Se} and the descendants, \isotope[64]{As} and \isotope[63]{Ge}. The $\gamma$-ray spectra associated with the three corresponding implants are presented in figures \ref{fig:se64_gamma_spectrum}, \ref{fig:as64_gamma_spectrum} and \ref{fig:ge63_gamma_spectrum}.
	
	%%%%%%%%%%%% Fig 3 ID plot and decaychain %%%%%%%%%%%%

	%%%%%%% Subsubsection gammas %%%%%%%%%%%%%%%%%%%
	%****************************************************
	\subsubsection{Beta-delayed gamma radiation}
	%****************************************************
	
\begin{figure}
		\includegraphics[width=8.6cm]{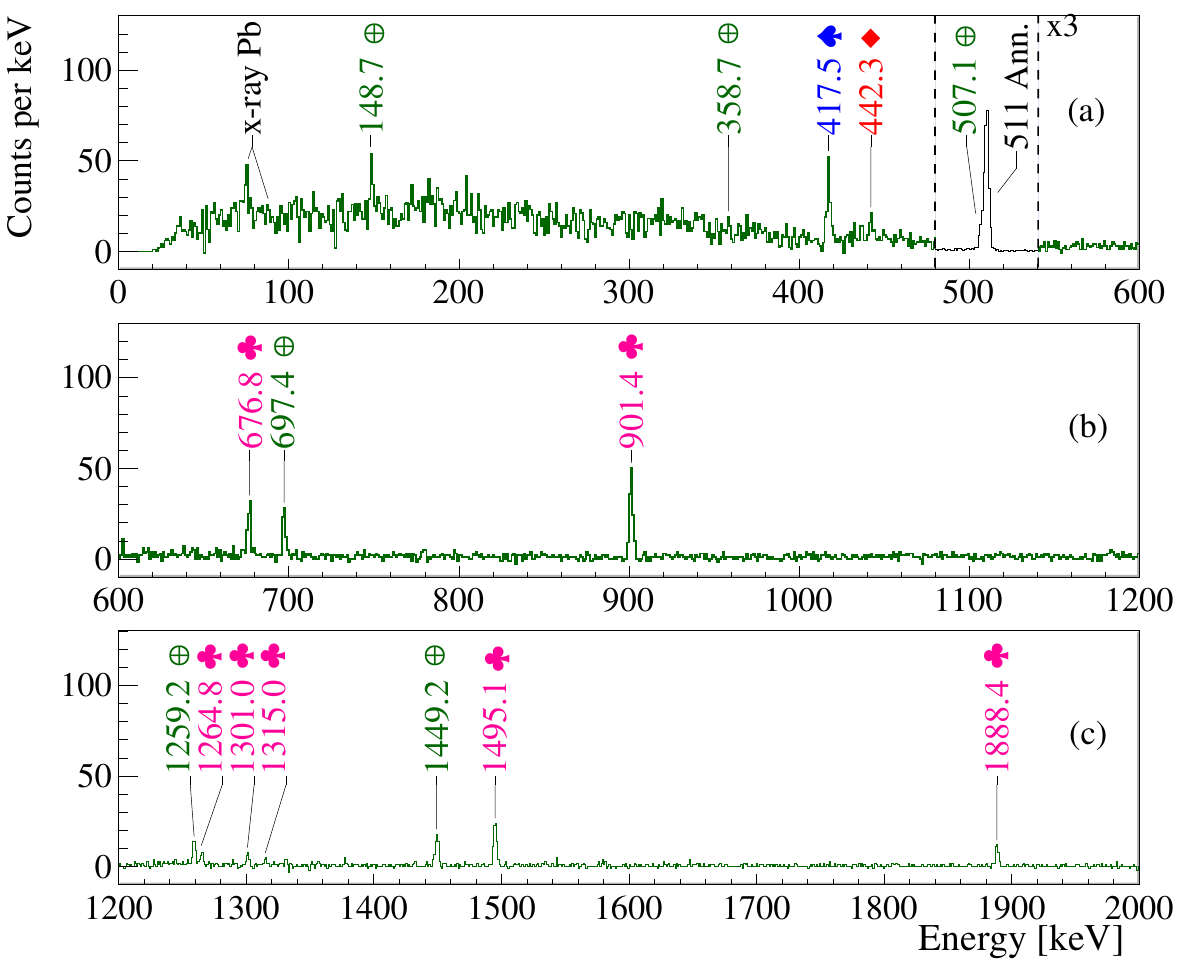}
		
		\caption{ 
			The figure shows the  $\gamma$-ray spectrum in time correlation with implants of \isotope[64]{Se} nuclei. The $\gamma$-rays assigned to  \isotope[64]{Se} decay in \isotope[64]{As} are shown in green and marked with $\oplus$, the $\gamma$-ray 
			assigned to \isotope[64]{Se} $\beta$-delayed p decay in \isotope[63]{Ge} is  shown in blue and marked with $\spadesuit$, the $\gamma$-rays assigned to \isotope[64]{As} decay in \isotope[64]{Ge} are shown in pink and marked with $\clubsuit$ and the $\gamma$-ray assigned to \isotope[63]{Ge} decay in \isotope[63]{Ga} is shown in red with $\blacklozenge$. The annihilation radiation and x-rays are shown in black. 
			The $511\mathrm{~keV}$ annihilation peak shows a low energy shoulder corresponding to a $\gamma$-ray in 
			the decay of \isotope[64]{Se}, see text.
			\label{fig:se64_gamma_spectrum}
		}
	\end{figure}

	Gamma-ray spectra were obtained using the correlations between each implantation event of the nucleus of interest, happening in one pixel, with all the beta-delayed $\gamma$-rays detected in EURICA, in prompt coincidence with the WAS3ABi decay signal happening in the same pixel or in an adjacent pixel during the correlation-time. 
	The correlation-time  was 7 times the decay half-life of the nucleus of interest. 
	This methodology includes a substantial portion of random events that have to be subtracted. 
	In order to do so, a similar spectrum was constructed using the same time-interval but in the backwards direction. 
	Namely, taking all the ``decay'' $\gamma$ events happening before the implantation. 
	The $\gamma$ spectra were  constructed in addbackmode. 
	The EURICA array was calibrated using standard sources of \isotope[152]{Eu} and \isotope[133]{Ba} together with Monte Carlo simulations at higher energies. 
	Figure \ref{fig:se64_gamma_spectrum}  shows the spectrum obtained in time correlation with the identified \isotope[64]{Se} implants, following backwards time correlation subtraction. In green, we show the $\gamma$-lines that are observed in time correlation with \isotope[64]{Se} implants but not with \isotope[64]{As} or 
	\isotope[63]{Ge}. They are associated with the beta decay of \isotope[64]{Se}. 
	
	%%%%%%%%%%%%%%% fig 4 Gamma spectrum 64Se %%%%%%%%%%%
	
\begin{figure}
		\includegraphics[width=8.6cm]{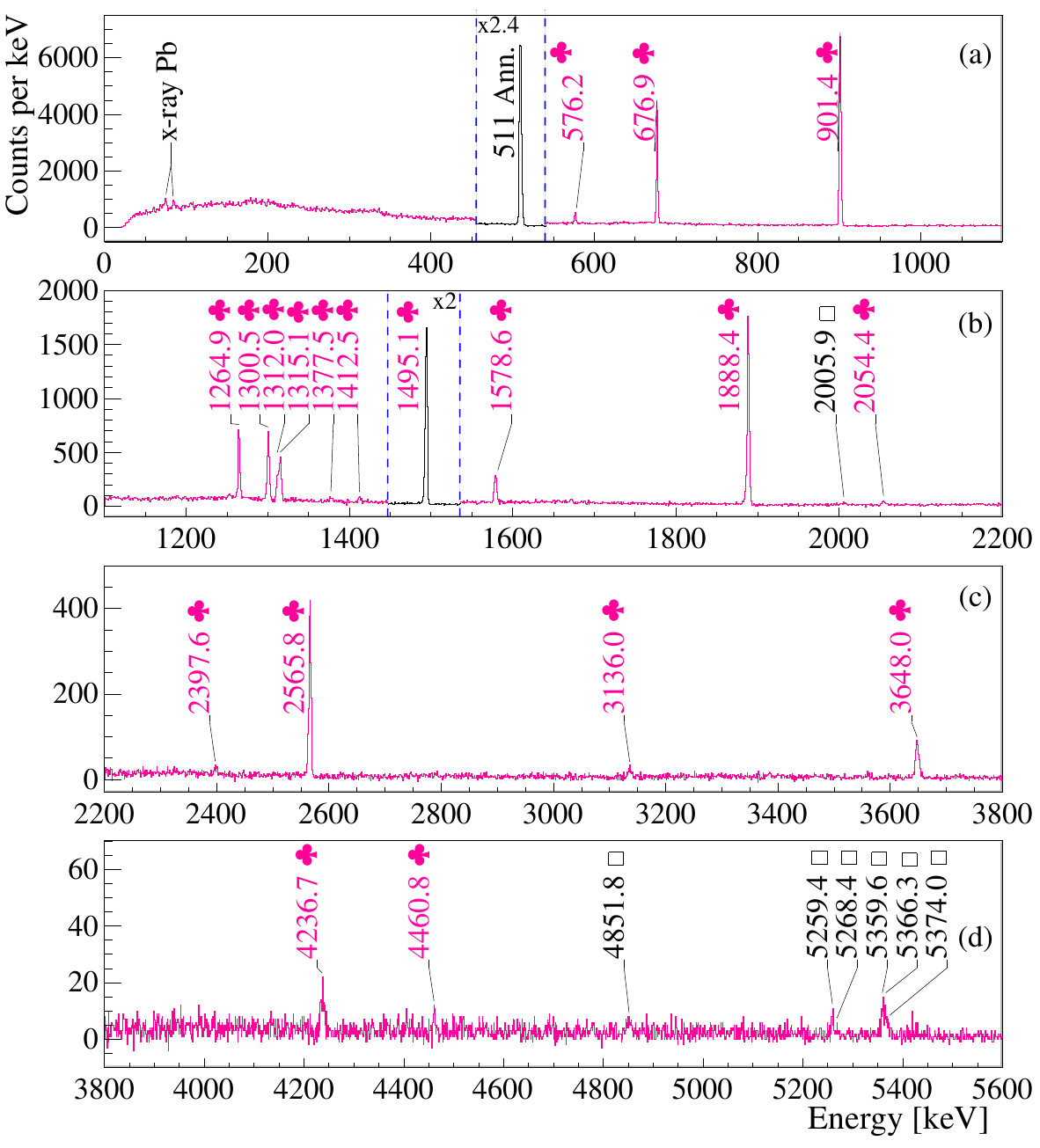}
		\caption{
			The figure shows the  $\gamma$-ray spectrum in time correlation with implants of \isotope[64]{As} nuclei within a $440\mathrm{~ms}$ implant-decay correlation window. Unplaced $\gamma$-rays are shown in black and marked with $\Box$, together with the Lead x-rays and the positron annihilation peak. The $\gamma$-rays assigned to \isotope[64]{As} $\beta$ decay are shown in pink and marked with $\clubsuit$.    
			\label{fig:as64_gamma_spectrum}}
	\end{figure}

	When we put the identification condition on \isotope[64]{As} implants, 
	we expect to observe $\gamma$-lines in \isotope[64]{Ge} populated  directly in $\beta$ decay, or $\gamma$-lines in \isotope[63]{Ga} populated by $\beta$-delayed proton-decay. 
	This is shown in Fig. \ref{fig:as64_gamma_spectrum}, where $\gamma$-lines belonging to \isotope[64]{Ge} are shown in pink.
	
	%%%%%%%%%%%%%%% fig 5 Gamma spectrum 64As %%%%%%%%%%%
	
	Finally, the $\gamma$-lines in \isotope[63]{Ga} are marked in red in Fig. \ref{fig:ge63_gamma_spectrum}, with the identification condition for the correlations set on the implantation of \isotope[63]{Ge}. No beta-delayed $\gamma$ radiation was known in any of these three decays prior to this work. 
	
\begin{figure}
		\includegraphics[width=8.6cm]{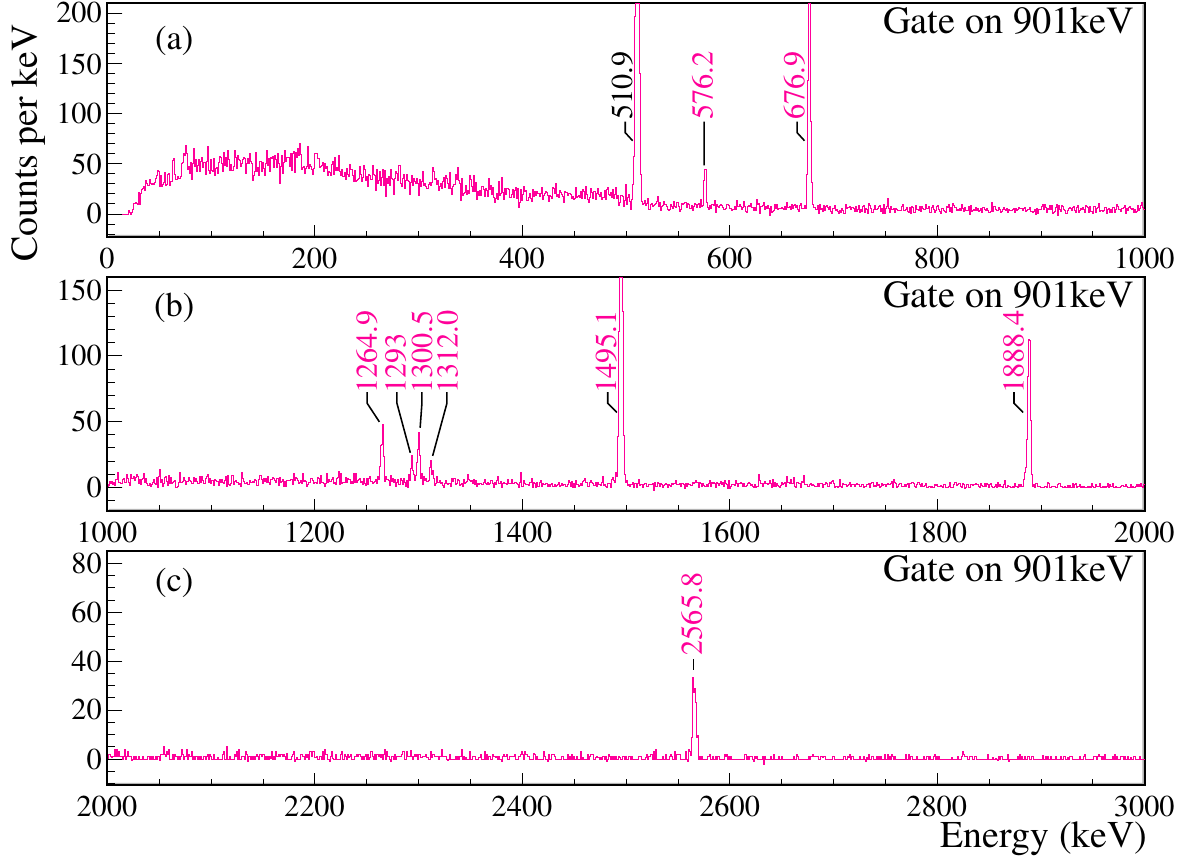}
		
		\caption{
			$\gamma$-$\gamma$ coincidence spectrum gated on the $901\mathrm{~keV}$ $\gamma$-ray for \isotope[64]{As} decay within a $440\mathrm{~ms}$ implant-decay correlation window.   
			\label{fig:as64_gg_gate901}}
	\end{figure}

\begin{figure}
		\includegraphics[width=8.6cm]{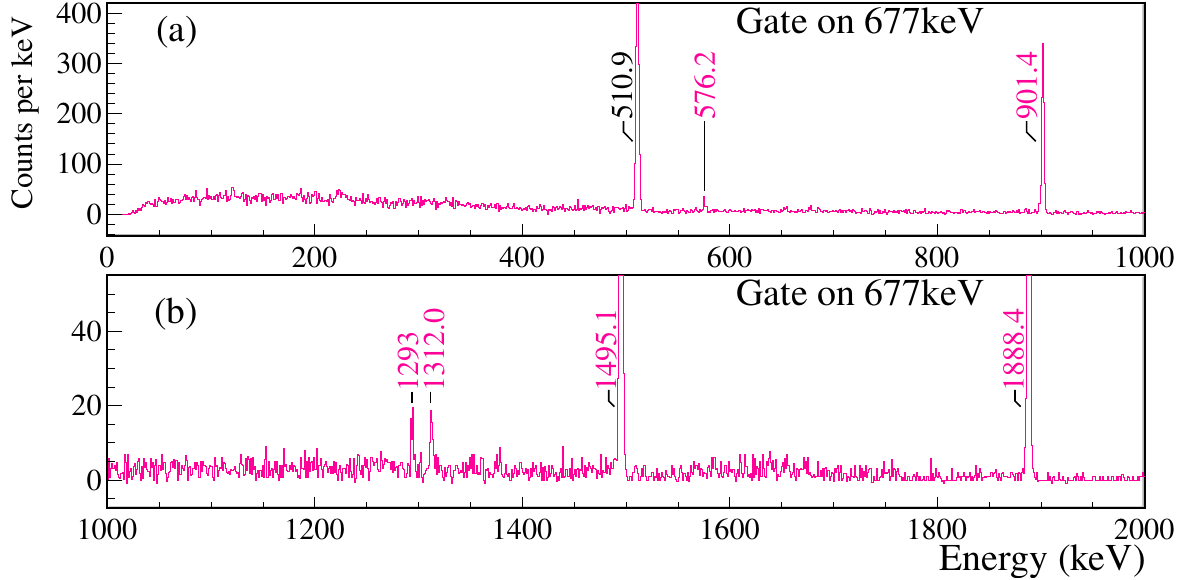}
		\caption{
			$\gamma$-$\gamma$ coincidence spectrum gated on the $677\mathrm{~keV}$ for \isotope[64]{As} decay within a $440\mathrm{~ms}$ implant-decay correlation window.   
			\label{fig:as64_gg_gate677}}
	\end{figure}

\begin{figure}
		\includegraphics[width=8.6cm]{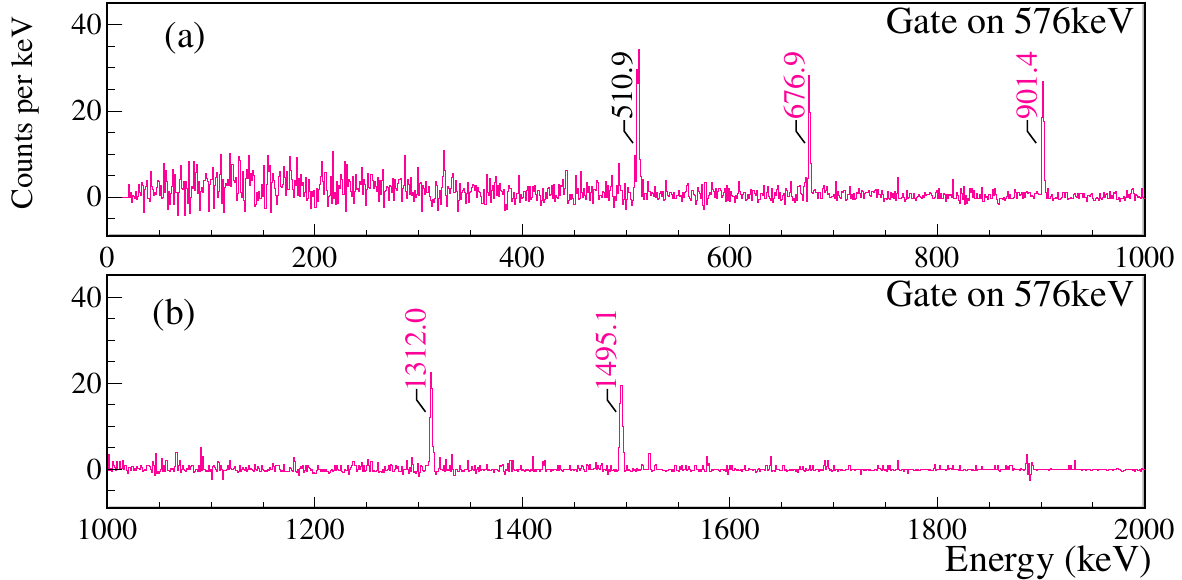}
		\caption{
			$\gamma$-$\gamma$ coincidence spectrum gated on the $576\mathrm{~keV}$ for \isotope[64]{As} decay within a $440\mathrm{~ms}$ implant-decay correlation window.   
			\label{fig:as64_gg_gate576}}
	\end{figure}

\begin{figure}
		\includegraphics[width=8.6cm]{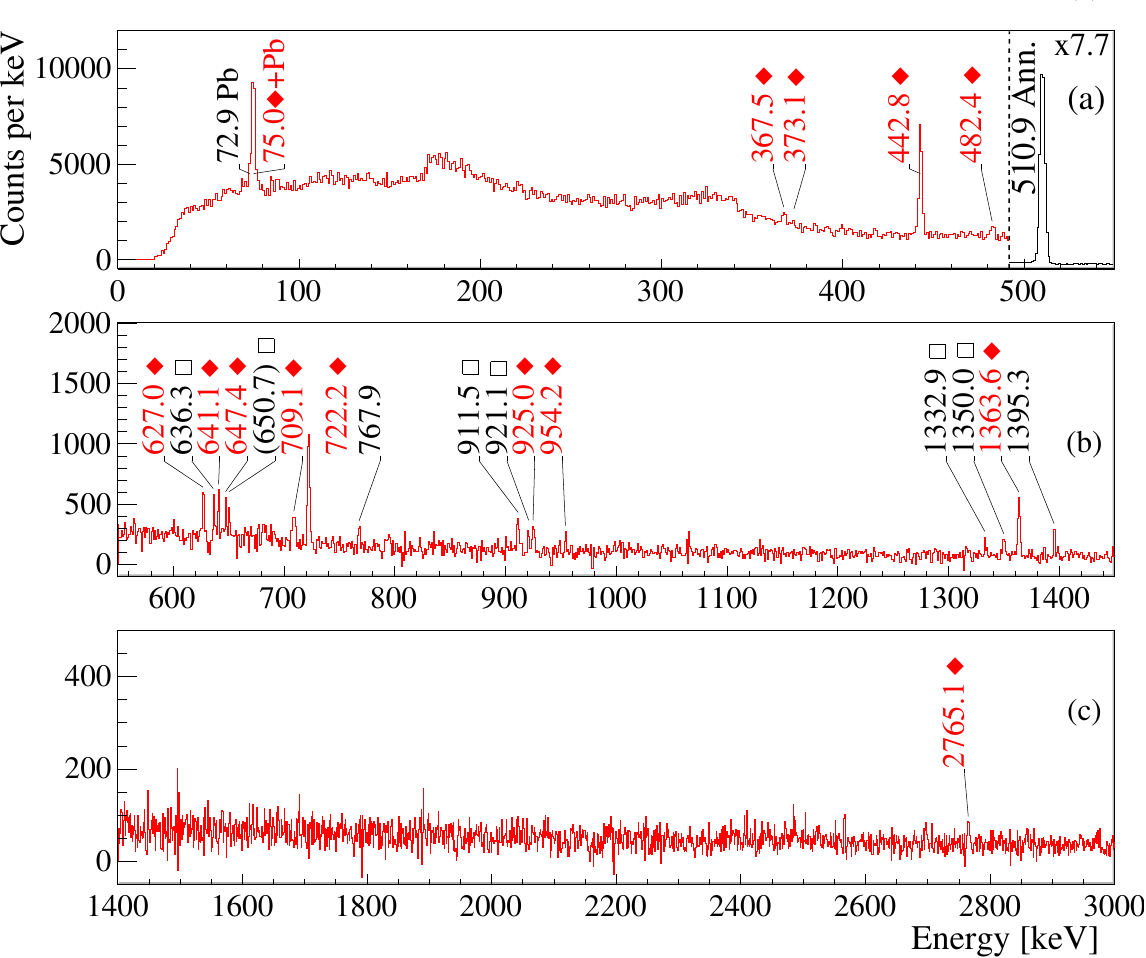}
		\caption{
			The figure shows the  $\gamma$-ray spectrum in time correlation with implants of \isotope[63]{Ge} nuclei  within a $1050\mathrm{~ms}$ implantation-decay correlation window. Unplaced $\gamma$-rays, the Lead x-rays and positron annihilation peak are all shown in black and marked with $\Box$. The $\gamma$-rays identified as belonging to 	\isotope[63]{Ge} beta decay are shown in red and marked with {\color{red} $\blacklozenge$}.
			\label{fig:ge63_gamma_spectrum}
		}
	\end{figure}

	Following the same procedure we constructed the $\gamma$-$\gamma$ matrix for \isotope[64]{As} decay to the N=Z nucleus \isotope[64]{Ge}. In Figs. \ref{fig:as64_gg_gate901}, \ref{fig:as64_gg_gate677} and \ref{fig:as64_gg_gate576}, we show the $\gamma$-ray spectra in coincidence with the $901\mathrm{~keV}$, $677\mathrm{~keV}$ and $576\mathrm{~keV}$ $\gamma$-rays, known from previous studies \cite{Ennis1991,Farnea2003}.

	%%%%%%%%%%%%%%% fig 6 Gamma spectrum 63Ge %%%%%%%%%%%
	
	%%%%%%%%%%%%  subsubsection protons %%%%%%
	%****************************************************
	\subsubsection{Beta radiation and beta-delayed proton emission}
	\label{sec:betas and protons section}
	%****************************************************
	
\begin{figure}
		\includegraphics[width=8.6cm]{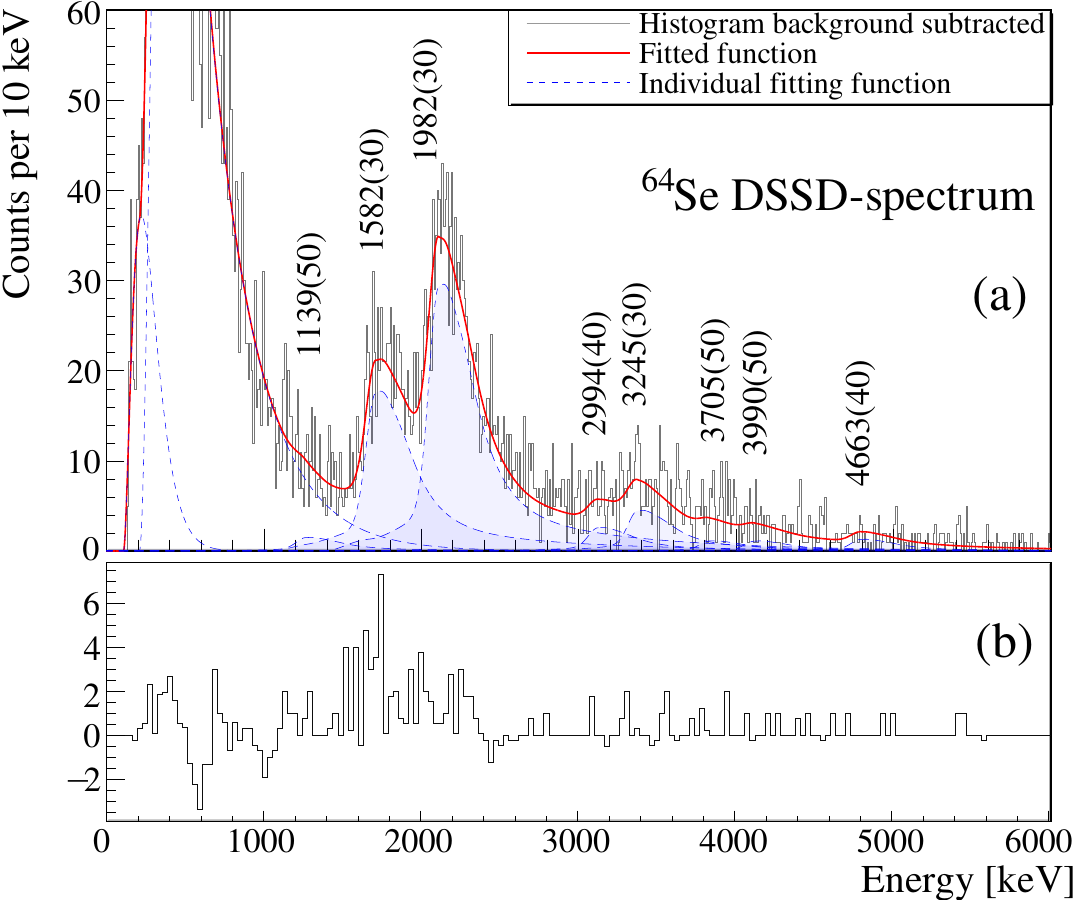}
		\caption{
			In the upper half of the figure we see the DSSSD spectrum correlated in time with implants of \isotope[64]{Se}. The implantation-decay forward correlation time window was $170\mathrm{~ms}$. The backwards correlated spectrum within the same time interval has been subtracted. Proton peaks are filled in light blue in the fit, while the beta contribution is the large peak at low  energy. The peaks are labelled with the proton energies, obtained as a sum of the fitted energy minus the DSSSD shift of $130\mathrm{~keV}$, see text. The peak at very low energy is attributed to electronic noise. In the lower half of the figure the same DSSSD spectrum is shown, now gated in addition by the $417.5\mathrm{~keV}$ $\gamma$-ray in \isotope[63]{Ge}. Some evidence of a coincidence with the %1.7 MeV 
			1.58 MeV proton peak can be observed.
			\label{fig:se64_pspectrum}
		}
	\end{figure}

\begin{figure}
		\includegraphics[width=8.6cm]{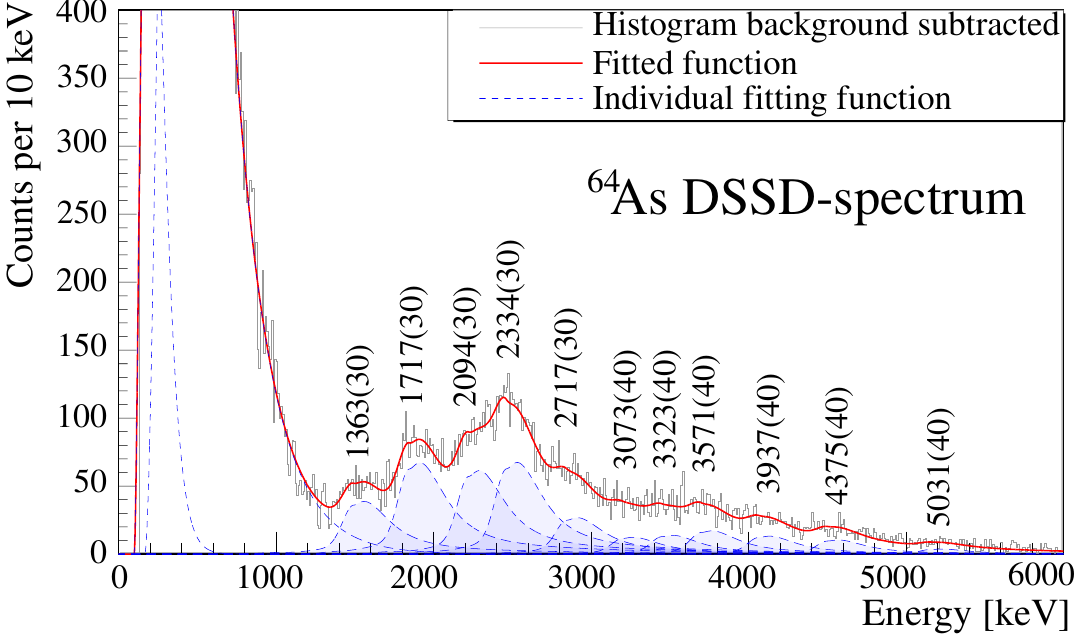}
		\caption{
			The beta-delayed proton spectrum for \isotope[64]{As} decay.  The implantation-decay forward correlation time window was $440\mathrm{~ms}$. The backwards correlated spectrum within the same time interval has been subtracted. 		Proton peaks are filled in light blue, while the beta contribution is shown to the left.
			The peaks are labelled with the proton energies, obtained as a sum of the fitted energy minus the DSSSD shift of $130\mathrm{~keV}$, see text.  The peak at very low energy is attributed to electronic noise.
			\label{fig:as64_pspectrum}
		}
	\end{figure}

	The spectra of beta particles and protons were obtained using the implantation-decay correlations in WAS3ABi.
	As in the case of the $\gamma$-rays, the correlation time was adjusted using a common criterion of a time window chosen equal to seven times the half-life of the parent nucleus. The results are  shown in figures  \ref{fig:se64_pspectrum} and  \ref{fig:as64_pspectrum} for \isotope[64]{Se} and \isotope[64]{As} decay. The lowest energy parts of the spectra are dominated by the partial deposition of the $\beta$ particle energy in the detector. Above approximately 1 MeV one can see the peaks corresponding to proton absorption in the detector.  The energy calibration of WAS3ABi was carried out strip by strip  
	using a \isotope[207]{Bi} source, placed successively in front of each DSSSD strip.

	%%%%%%%%% fig 7 64Se protons %%%%%%%%%%%%%%%%%%%%
	
	%%%%%%%%% fig 8 64As protons %%%%%%%%%%%%%%%%%%%%

	Since the proton is emitted almost simultaneously with the beta particle the energy recorded is their sum. As a result there is a shift in the energies of the proton peaks in the spectra with respect to the calibration as well as a broadening in energy. 
	To determine the energy shift and the shape of the proton peaks, we studied other proton-emitting nuclei, with known proton energies, produced during our experiment, that were also implanted in WAS3ABi namely \isotope[61]{Ge}, \isotope[57]{Zn} and \isotope[65]{Se} \cite{aguilera_jorquera_study_2020,rubio_beta_2019}. 
	We observed an energy shift of $ \Delta E\!=\!130 (27) \mathrm{~keV} $ that is subtracted from the observed energies of the proton peaks in \isotope[64]{Se} and \isotope[64]{As} decays. The resulting proton energies are indicated in Figs. \ref{fig:se64_pspectrum} and \ref{fig:as64_pspectrum}.
	
	%***************************************************
	\subsubsection{The decay scheme of \isotope[64]{Se} and the ground state mass of \isotope[64]{As}}
	\label{sec:levelsch_64se}
	%**************************************************
	
\begin{figure}
		\includegraphics[width=8.6cm]{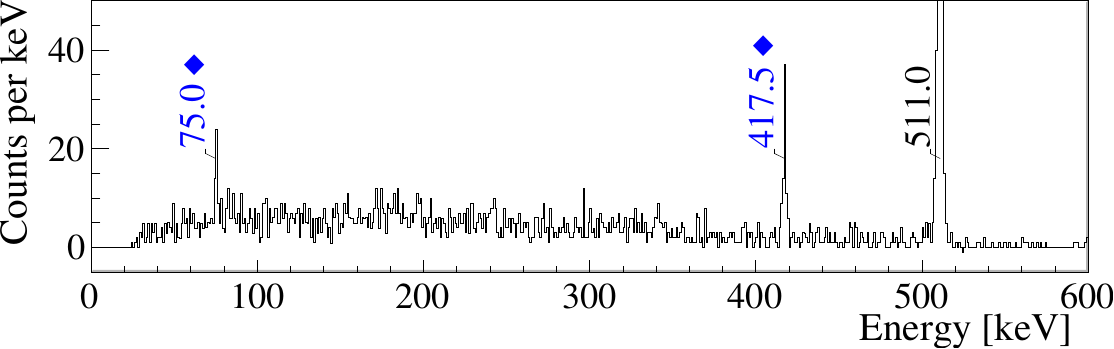}
		\caption{
			The $\gamma$-ray spectrum for \isotope[64]{Se} decay, selecting the proton events in the DSSSD. Correlations are done between the \isotope[64]{Se} implantations and the decay protons within $170\mathrm{~ms}$.
			$\gamma$-rays attributed to transitions in the daughter nucleus \isotope[63]{Ge} are labelled in blue and marked with $\blacklozenge$, see text.
			\label{fig:se64_pgated_gammas} 
		}
	\end{figure}

\begin{figure}
		\includegraphics[width=8.6cm]{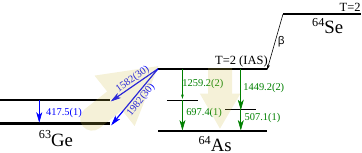}
		\caption{
			Diagram of the decay of  \isotope[64]{Se}, illustrating the method used to obtain the mass excess of the ground state of \isotope[64]{As}. The energy of the protons is obtained from the proton peaks shown in Fig. \ref{fig:se64_pspectrum}, after a $130(27)\mathrm{~keV}$ correction, see text. The \isotope[63]{Ge} ground state mass excess $-46920(40)\mathrm{~keV}$ plus the proton energy minus the excitation energy of the IAS in \isotope[64]{As} allows us to determine the mass excess for \isotope[64]{As} g.s.  as $-39588(50)\mathrm{~keV}$. \label{fig:MassScheme_63Ge-64As}}
	\end{figure}

	As mentioned above nothing was known previously about the excited states of \isotope[64]{As}, the daughter nucleus of \isotope[64]{Se}. We have clearly identified $\gamma$ transitions which belong to the decay of \isotope[64]{Se} to \isotope[64]{As}. The proton binding energy in \isotope[64]{As} is very small, $-100(200)\mathrm{~keV}$, according to the latest AME2020 \cite{wang_ame_2021}.
	Consequently, the proton spectrum shown in Fig. \ref{fig:se64_pspectrum} should be dominated by the delayed protons after the $\beta$-decay of \isotope[64]{Se}. Furthermore, no information exists on the excited states of \isotope[63]{Ge}, the final nucleus populated by proton emission, but a $\gamma$-ray with energy 
	$442.8\mathrm{~keV}$ is known in the mirror nucleus \isotope[63]{Ga} \cite{Weiszflog2001,rudolph_2021}.
	The energy difference between the two largest peaks in the proton spectrum is around $400\mathrm{~keV}$. 
	In addition, there is  a  $\gamma$-ray peak of $417.5\mathrm{~keV}$ with the condition on \isotope[64]{Se} implantations and in coincidence with the protons as shown in Fig. \ref{fig:se64_pgated_gammas}, which we assign to the de-excitation of a state  in \isotope[63]{Ge} of this energy (see Fig. \ref{fig:MassScheme_63Ge-64As}, left side). It should be noted that a $\gamma$ line with a similar energy to the observed $417.5(1)\mathrm{~keV}$, namely $419.0(7)\mathrm{~keV}$, has been observed in a two proton knock out reaction at MSU \cite{henry_thesis_2015} and assigned to \isotope[63]{Ge}. 
	We believe that the two proton peaks with energies $1582(30)\mathrm{~keV}$ and $1982(30)\mathrm{ ~keV}$ 
	correspond to the de-excitation of a strongly populated state in  \isotope[64]{As} to the $417.5\mathrm{~keV}$ excited state and the ground state in \isotope[63]{Ge}, as shown in Fig. \ref{fig:MassScheme_63Ge-64As}.
	This is confirmed by the fact that putting  a condition on the $417.5\mathrm{~keV}$ $\gamma$ line and looking at the proton spectrum, the higher peak of the doublet, with energy $1982\mathrm{~keV}$, disappears while there is a coincidence with the first peak of the doublet (see Fig. \ref{fig:se64_pspectrum} lower half). 
	Another gamma line with energy $75.0\mathrm{~keV}$ has been identified in the $\beta$-delayed proton spectrum of \isotope[64]{Se} shown in Fig. \ref{fig:se64_pgated_gammas}. Based on the existence of a level with similar energy in the mirror nucleus, we have tentatively assigned it as the de-excitation of the first excited state in \isotope[63]{Ge}. We note here that our experiment is not sensitive enough to assess if part of the $1982\mathrm{~keV}$ peak includes a proton transition to this level.
	
	%%%%%%%%%%% Fig 10 gammas 64Se gated on protones
	
	%%%%%%%%% fig 9 IMME Mass acrobatics %%%%%%%%%%%%%%%%%%%%
	
	One expects the decay of \isotope[64]{Se} to be dominated by the very fast Fermi (F) transition, and consequently we assign the excited state where the protons originate to be the IAS of \isotope[64]{Se}
	in \isotope[64]{As}. This is further corroborated by the fact that the excitation energy of this level can be estimated as illustrated in Fig.\ref{fig:MassScheme_63Ge-64As}, taking into account the proton energy of the $1582 \mathrm{~keV}$ peak, the gamma ray energy of $417.5\mathrm{~keV}$ and the atomic masses of \isotope[63]{Ge} and \isotope[64]{As} from \cite{wang_ame_2021}, namely $M(\isotope[64]{As})^\#\!=\!-39530^\#(200^\#)\mathrm{~keV}$ and $M(\isotope[63]{Ge})\!=\!-46920(40)\mathrm{~keV}$, where $\#$ indicates values from systematics. We note that in the calculation we have considered  only the $1582\mathrm{~keV}$ proton peak because the $1982\mathrm{~keV}$ peak may have some component to the possible first excited state at $75\mathrm{~keV}$.  The resulting value of $1898(200^\#)\mathrm{~keV}$ lies very close to the energy of the IAS in the mirror nucleus \isotope[64]{Ga}, namely $1923\mathrm{~keV}$ \cite{Diel2019}. 
	
	The proton (isospin $T\!=\!1/2$) decay of the $T\!=\!2$ IAS to the final nucleus \isotope[63]{Ge} low-lying states ($T\!=\!1/2$) is isospin forbidden and consequently one expects this decay to be retarded and thus in competition with $\gamma$ decay. 
	This has been the observation in the decay of other $T\!=\!2$ nuclei with lower masses. See for example the decays of \isotope[60]{Ge} and \isotope[56]{Zn} \cite{orrigo2021, rubio_beta_2014}.

	Looking at the $\gamma$ lines associated with the decay of \isotope[64]{Se} in Fig. \ref{fig:se64_gamma_spectrum}, we see two strong transitions, 697.4 and $1259.2\mathrm{~keV}$, summing to $1956.6\mathrm{~keV}$, very close to the estimated energy for the IAS. We assign these two $\gamma$ lines to the de-excitation of the IAS, which thus becomes $1956.5(2)\mathrm{~keV}$. With this assumption we can now fix the mass of the \isotope[64]{As} g.s. (ground state) as,
	\begin{widetext}
		\begin{align}
			M(\isotope[64]{As},\mathrm{g.s.}) &= M(\isotope[63]{Ge}) + M(p) + 1582(30)\mathrm{~keV} +417.5(1)\mathrm{~keV} -1956.5(2) \mathrm{~keV} = -39588(50) \mathrm{~keV}
		\end{align}
	\end{widetext}
	
	This result is in very good agreement with a recent direct mass measurement of the  \isotope[64]{As} ground state, $-39710(110) \mathrm{~keV}$, carried out at the Lanzhou facility based on 6 events 
	\cite{zhou_mass_2023}. 
	
\begin{figure*}
		\includegraphics[width=17.2cm]{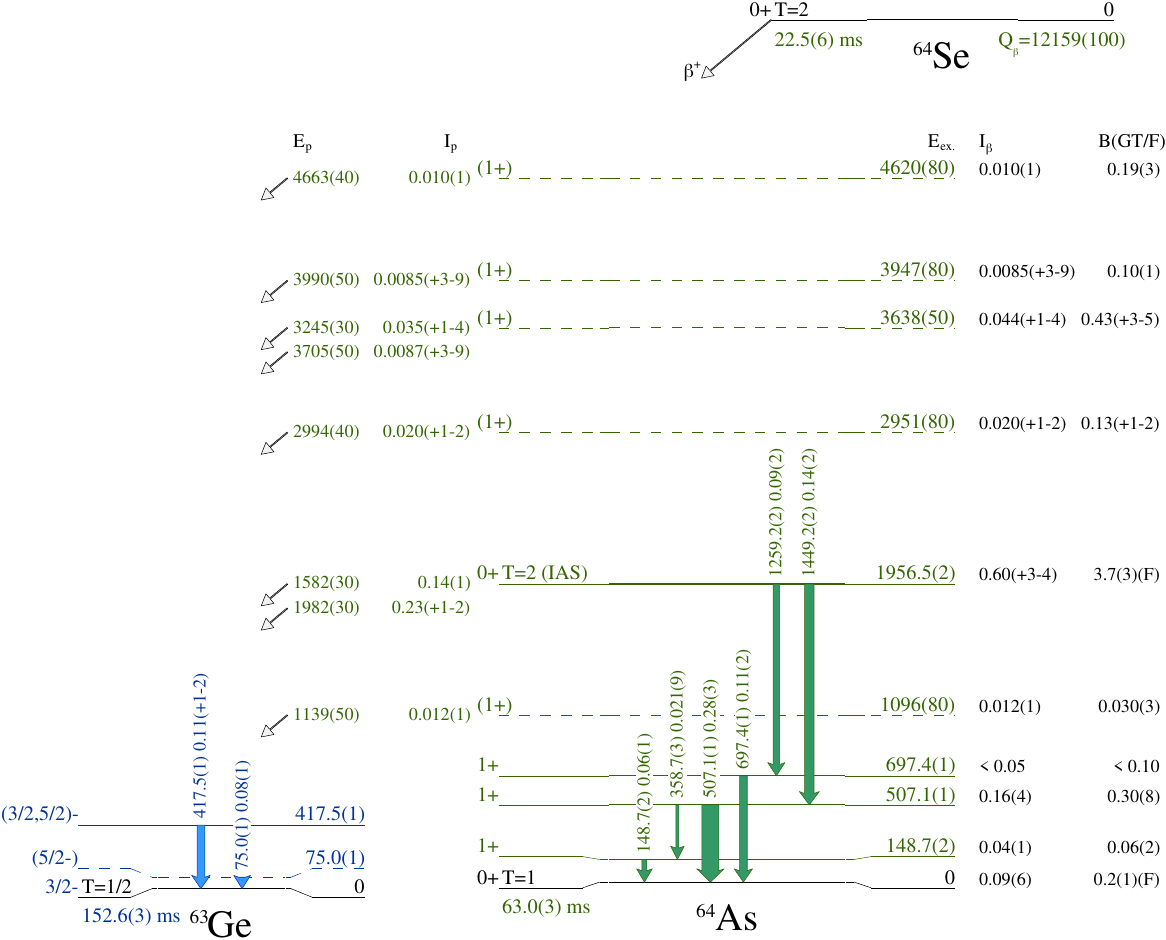}
		\caption{		
			Decay scheme for \isotope[64]{Se}, half-life and B(GT/F) values are based on the present experiment.  Transition intensities are given per decay.  The mass excess value for \isotope[64]{As} was deduced using the \isotope[63]{Ge} mass excess from \cite{wang_ame_2021} and our data (see text). The  $ Q_\beta $ value for \isotope[64]{Se} was calculated as the difference between the experimental mass excess for \isotope[64]{As} deduced from our results and the extrapolated value for the \isotope[64]{Se} ground state mass excess using the IMME equation (see text). 
			Half-lives of \isotope[63]{Ge} and \isotope[64]{As} are based on the present experiment. Transition strength for the ground state is calculated  assuming a Fermi transition, although it is an isospin forbidden decay.
			\label{fig:se64_levsch} }
	\end{figure*}

	Another strong $\gamma$ peak associated with the decay of \isotope[64]{Se} lies at $1449.2\mathrm{~keV}$
	having a difference of $507.1\mathrm{~keV}$ with the energy of the IAS. In Fig. \ref{fig:se64_gamma_spectrum} one can see clearly that this peak exists although it is not resolved from the strong $511\mathrm{~keV}$ annihilation peak. 
	Moreover, the existence of a level of $507.1\mathrm{~keV}$ is further confirmed by a weak coincidence between the $1449.2(2)\mathrm{~keV}$ with the $358.7(3)\mathrm{~keV}$ and $148.7(2)\mathrm{~keV}$ transitions, a cascade in parallel with the $507.1(1)\mathrm{~keV}$ transition. The full level scheme summarising all these facts can be seen in Fig. \ref{fig:se64_levsch} where we have also included levels corresponding to the states populated by the other weak proton peaks that were observed. We have assumed that these transitions proceed to the  ground state except when the difference between two peaks coincided with the energy of the 417 keV  excited state in \isotope[63]{Ge}. The direct feeding to each level was calculated as the difference between the total $\gamma$ intensity populating  the level   and the sum of the $\gamma$-ray and/or proton intensity de-exciting the level. The ground state to ground state feeding was estimated as the difference between the $\beta$ branching and the $\gamma$ feeding to the ground state. The properties of the levels are summarised in Table \ref{Tab:intensities_64Se_decay}.

	%%%%%%%%% fig 18   64Se Decay scheme  %%%%%%%%%%%%%%%%%%%%

	%\begin{widetext}
	%\begin{turnpage}
	\begin{table*}
		\caption{ The table shows the properties of the levels in  \isotope[64]{As} populated in the $\beta$ decay of \isotope[64]{Se}.
			Column 1 indicates the energies of the levels in \isotope[64]{As}. Column  2,  the assigned spin-parity, columns 3 and 4 the $\gamma$ de-excitation energy and intensity per decay, columns 5 and 6 the proton energy and intensity, column 7 the $\beta$ feeding to the level, and column 8 the B(GT) or B(F). 
			Transition strength for the ground state is calculated  assuming a Fermi transition, although it is an isospin forbidden decay.
			The properties of the IAS are marked with numbers in bold.
			\label{Tab:intensities_64Se_decay}}
		
		\begin{ruledtabular}
			\begin{tabular}{llllllll}
				
				$E_{ex}$ keV&  $J_\pi$  &   $E_\gamma$ keV  &   $I_\gamma$  &   $E_p$ keV      &   $I_p$       &  Feeding      &  $B(GT/F)$    \\\hline\hline
				0.0         &  $0^+$    &               &               &               &               &  0.09(6)      &  0.2(1) (F)   \\\hline
				148.7(2)    &  $1^+$    &   148.7(2)    &   0.06(1)     &               &               &  0.04(1)      &  0.06(2)      \\\hline
				507.1(1)    &  $1^+$    &   358.7(3)    &   0.021(9)    &               &               &  0.16(4)      &  0.30(8)      \\
				&           &   507.1(1)    &   0.28(3)     &               &               &               &               \\\hline
				697.4(1)    &  $1^+$    &   697.4(1)    &   0.11(2)     &               &               &  $<0.05$      &  $<0.10$      \\\hline
				1096(80)    &  $(1^+)$    &               &               &   1139(50)    &  0.012(1)     &  0.012(1)     &  0.030(3)     \\\hline
				\bf 1956.5(2) IAS & $\mathbf{0^+\, , T\!=\!2}$ 
				& \bf 1259.2(2) & \bf  0.09(2)  &  \bf 1582(30) & \bf 0.14(1)   & \bf 0.60(+3-4)&  \bf 3.7(3) (F)   \\
				&           & \bf 1449.2(2) & \bf 0.14(2)   &  \bf 1982(30) & \bf 0.23(+1-2)&               &               \\\hline
				2951(80)    & $(1^+)$     &               &               &   2994(40)    &   0.020(+1-2) &  0.020(+1-2)  &  0.13(+1-2)  \\\hline
				3638(50)    & $(1^+)$     &               &               &   3245(30)    &   0.035(+1-4) &  0.044(+1-4)  &  0.43(+3-5)  \\
				&           &               &               &   3705(50)    &   0.0087(+3-9)&               &               \\\hline
				3947(80)    & $(1^+)$     &               &               &   3990(50)    &   0.0085(+3-9)&  0.0085(+3-9) &  0.10(1)      \\\hline
				4620(80)    & $(1^+)$     &               &               &   4663(40)    &   0.010(1)    &  0.010(1)     &  0.19(3)      \\   
			\end{tabular}
		\end{ruledtabular}
	\end{table*}
	%\end{turnpage}
	%\end{widetext}

	%***************************************************
	\subsubsection{Decay schemes of \isotope[64]{As} and
		\isotope[63]{Ge} }
	\label{sec:levelsch_64as_63ge}
	%**************************************************
	
\begin{figure*}
		\includegraphics[width=17.2cm]{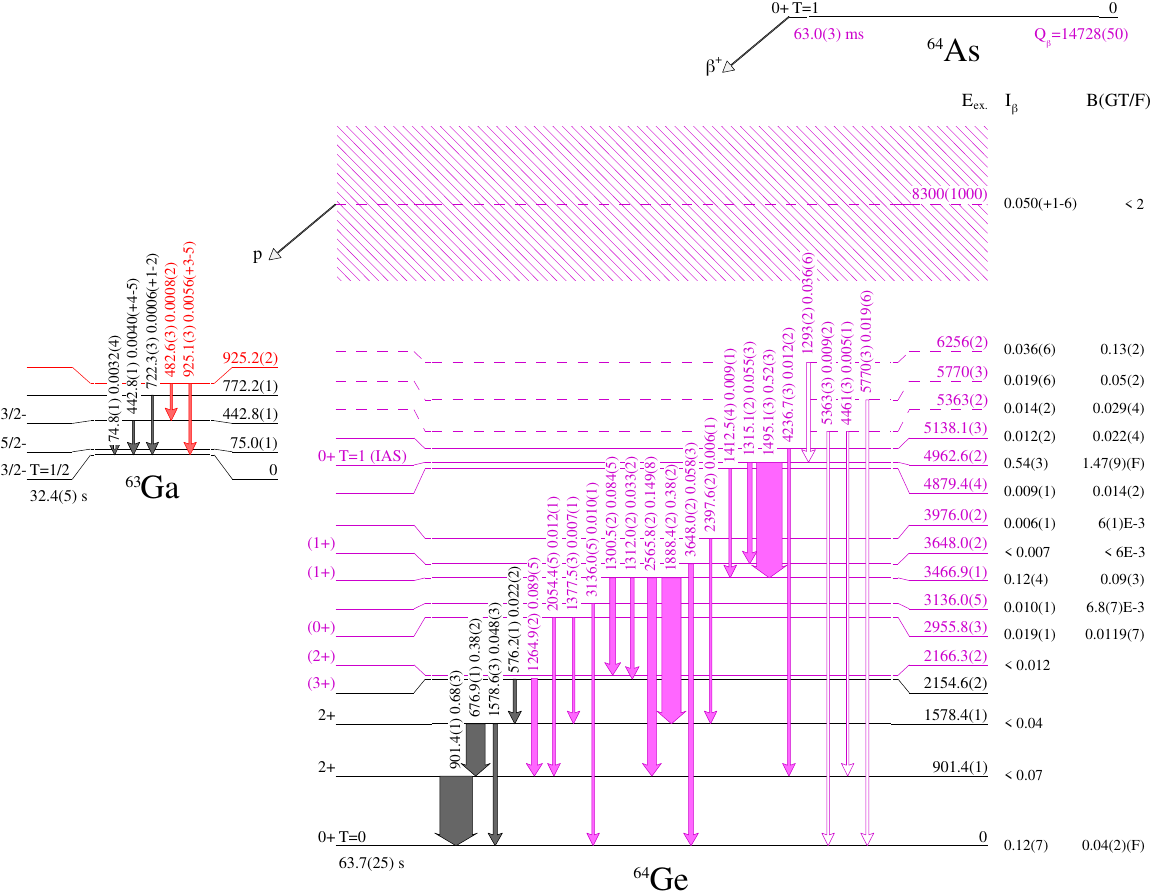}
		\caption{    
			Decay scheme for \isotope[64]{As}, half-life and B(GT/F) values are based on the present experiment.  Transition intensities are given per decay. The $Q_\beta$ value was calculated using the mass deduced in this work for \isotope[64]{As} (see text) and the \isotope[64]{Ge} mass from AME2020 \cite{wang_ame_2021}.         The transitions shown in colour represent $\gamma$-rays and states observed for the first time. Tentative levels are drawn as dashed lines. Shaded area corresponds to the energy range covered by the observed proton spectrum (see Fig. \ref{fig:as64_pspectrum}).  Half-lives for \isotope[63]{Ga} and \isotope[64]{Ge} were taken from \cite{nubase2020}. Transition strength for the ground state is calculated  assuming a Fermi transition, although it is an isospin forbidden decay. For levels without spin-parity assignment, $1^+$ is assumed for the B(GT) calculations.
			\label{fig:as64_levsch} 
		}
	\end{figure*}

	As mentioned above, in order to identify the $\gamma$-rays emitted in the decay of \isotope[64]{Se} and construct the decay scheme, one must first construct the decay schemes of \isotope[64]{As} and \isotope[63]{Ge}. In the following we explain how we have analysed the decays of these isotopes and present the corresponding decay schemes.
	We shall begin with the decay of \isotope[64]{As}.
	The data were selected by placing the appropriate conditions to identify this nucleus on the ID plot shown in Fig. \ref{fig:pid_3001}, and looking at the correlated particles or $\gamma$ radiation within a time
	interval of $440\mathrm{~ms}$. Prior to the present work, excited states in \isotope[64]{Ge} had only been studied in-beam in reactions \cite{Ennis1991,Farnea2003}. The decays of three of the excited states in \isotope[64]{Ge} that were observed in these experiments, were also observed in the present work. They are shown in black in the decay scheme in Fig. \ref{fig:as64_levsch} together with the de-exciting $\gamma$ rays.

	%%%%%%%%% fig 11 64As decay scheme %%%%%%%%%%%%%%%%%%%%

	In the following, the decay scheme was constructed based on this information and the observed $\gamma$-$\gamma$ coincidences or matching energy differences. Some examples of coincidence gates are shown in 
	Figs. \ref{fig:as64_gg_gate901}, \ref{fig:as64_gg_gate677} and \ref{fig:as64_gg_gate576}. However, considering that \isotope[64]{As} has 0$^{+}$ ground state, one has to consider the possibility of populating  1$^{+}$ states 	in \isotope[64]{Ge} that decay only to the g.s. In fact there are some unidentified $\gamma$-rays (see Fig. \ref{fig:as64_gamma_spectrum}) which fulfil this condition. We have tentatively assigned the strong $5363(3)\mathrm{~keV}$  and $5770(3)\mathrm{~keV}$ peaks as transitions of this kind. Moreover, a possible transition de-exciting the $5363\mathrm{~keV}$ level with $4461\mathrm{~keV}$ energy could feed the 2$^{+}$
	level at $901.4\mathrm{~keV}$ energy. All these facts are summarised in the decay scheme shown in Fig. \ref{fig:as64_levsch} and in Table \ref{Tab:intensities_64As_decay}. The J$^{\pi}$ assignments will be discussed later in the discussion section which follows.

	%\begin{widetext}
	%\begin{turnpage}
	\begin{table*}
		\caption{ The table shows the properties of the levels in  \isotope[64]{Ge} populated in the $\beta$ decay of \isotope[64]{As}.
			Column 1 indicates the energies of the levels in \isotope[64]{Ge}. Column  2, the assigned spin-parity, columns 3 and 4 the  $\gamma$-ray energy and intensity per decay, columns 5 and 6 the proton energy and intensity, column 7 the $\beta$ feeding to the level, and column 8 the B(GT) or B(F). For levels without spin-parity assignment, $1^+$ is assumed for  the B(GT) calculations. Transition strength for the ground state is calculated  assuming a Fermi transition, although it is an isospin forbidden decay.
			The properties of the IAS are marked with numbers in bold.
			\label{Tab:intensities_64As_decay}}
		
		\begin{ruledtabular}
			\begin{tabular}{llllllll}
				
				$E_{ex}$ keV&  $J_\pi$  &  $E_\gamma$ keV  &  $I_\gamma$   &   $E_p$  keV &   $I_p$   &  Feeding      &  $B(GT/F)$        \\\hline\hline
				0.0         &  $0^+$    &               &               &           &           &  0.12(7)      &  0.04(2) (F)      \\\hline
				901.4(1)    &  $2^+$    &  901.4(1)     &  0.68(3)      &           &           &  $<0.07$      &  -                \\\hline
				1578.4(1)   &  $2^+$  &  676.9(1)     &  0.38(2)      &           &           &  $<0.04$      &  -                \\
				&           &  1578.6(3)    &  0.048(3)     &           &           &               &                   \\\hline
				2154.6(2)   &  $(3^+)$  &  576.2(1)     &  0.022(2)     &           &           &  -            &  -                \\\hline
				2166.3(2)   &  $(2^+)$  &  1264.9(2)    &  0.089(5)     &           &           &  $<0.012$     &  -                \\\hline
				2955.8(3)   &  $(0^+)$  &  2054.4(5)    &  0.012(1)     &           &           &  0.019(1)     &  0.0119(7)        \\
				&           &  1377.5(3)    &  0.007(1)     &           &           &               &                   \\\hline
				3136.0(5)   &           &  3136.0(5)    &  0.010(1)     &           &           &  0.010(1)     &  $0.0068(7)$       \\\hline
				3466.9(1)   &  $(1^+)$  &  1300.5(2)    &  0.084(5)     &           &           &  0.12(4)      &  0.09(3)          \\
				&           &  1312.0(2)    &  0.033(2)     &           &           &               &                   \\
				&           &  2565.8(2)    &  0.149(8)     &           &           &               &                   \\
				&           &  1888.4(2)    &  0.38(2)      &           &           &               &                   \\\hline
				3648.0(2)   &  $(1^+)$  &  3648.0(2)    &  0.058(3)     &           &           &  $<0.007$     &  $<0.006$         \\\hline
				3976.0(2)   &           &  2397.6(2)    &  $0.006(1)$   &           &           &  $0.006(1)$   &  $0.006(1)$       \\\hline
				4879.4(4)   &           &  1412.5(4)    &  $0.009(1)$   &           &           &  $0.009(1)$   &  0.014(2)         \\\hline
				\bf 4962.6(2) IAS & $\mathbf{0^+,\,  T\!=\!1}$ 
				& \bf 1315.1(2) &  \bf 0.055(3) &           &           & \bf 0.54(3)   & \bf 1.47(9) (F)   \\
				&           & \bf 1495.1(3) &  \bf 0.52(3)  &           &           &               &                   \\\hline
				5138.1(3)   &           &  4236.7(3)    &  0.012(2)     &           &           &  0.012(2)     &  0.022(4)         \\\hline
				5363(2)     &           &  5363(3)      &  0.009(2)     &           &           &  0.014(2)     &  0.029(4)         \\
				&           &  4461(3)      &  0.005(1)     &           &           &               &                   \\\hline
				5770(3)     &           &  5770(3)      &  0.019(6)     &           &           &  0.019(6)     &  0.05(2)          \\\hline
				6256(2)     &           &  1293(2)      &  0.036(6)     &           &           &  0.036(6)     &  0.13(2)          \\\hline
				P states    &           &               &               &  1363(30) &  0.0058(5)&  0.050(+1-6)  &  $<2$             \\
				$\sim 8.3(10)\times10^3$& &             &               &  1717(30) &  0.0100(8)&               &                   \\
				&           &               &               &  2094(30) &  0.0092(8)&               &                   \\
				&           &               &               &  2334(30) &  0.0101(9)&               &                   \\
				&           &               &               &  2717(30) &  0.0040(3)&               &                   \\
				&           &               &               &  3073(40) &  0.0018(2)&               &                   \\
				&           &               &               &  3323(40) &  0.0021(3)&               &                   \\
				&           &               &               &  3571(40) &  0.0026(3)&               &                   \\
				&           &               &               &  3937(40) &  0.0020(2)&               &                   \\
				&           &               &               &  4375(40) &  0.0015(1)&               &                   \\
				&           &               &               &  5031(40) &  0.0006(1)&               &                   \\
			\end{tabular}
		\end{ruledtabular}
	\end{table*}
	%\end{turnpage}
	%\end{widetext}

	As we have discussed earlier in section \ref{sec:betas and protons section}, there is a small amount of $\beta$-delayed proton emission in this decay. The spectrum is shown in Fig. \ref{fig:as64_pspectrum}.
	Unfortunately, the resolution was not good enough to clearly resolve all of the peaks. As shown in the figure one can have a reasonable fit to the data with a number of proposed proton transitions. However, this can only be seen as a good  approximation given the limited resolution. There may well be a number of other unresolved weak peaks in the spectrum.
	
\begin{figure}
		\includegraphics[width=8.6cm]{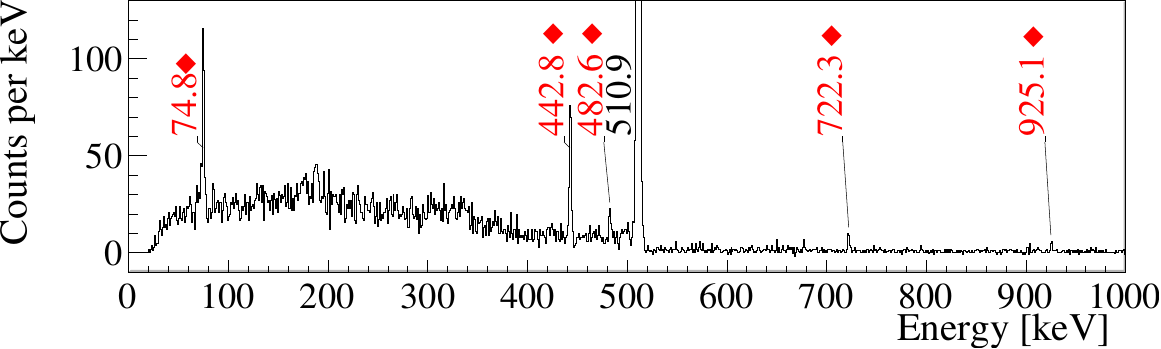}
		\caption{
			Gamma-ray spectrum observed in the decay of \isotope[64]{As}, selecting the proton events in the DSSSD, within a $440\mathrm{~ms}$ correlation window. The observed $\gamma$-rays are attributed to transitions in the daughter nucleus \isotope[63]{Ga}, see text. Please note that the energies quoted in this figure are obtained with these conditions and may differ slightly from the final adopted values reported in Table \ref{Tab:intensities_63Ge_decay} and Figure \ref{fig:ge63_levsch}. \label{fig:as64_pgated_gammaspectrum} 
		}
	\end{figure}

	We also observed coincidences between the protons and the $\gamma$ lines in the final \isotope[63]{Ga} nucleus, as shown in Fig. \ref{fig:as64_pgated_gammaspectrum}. From these observations we concluded that the beta-delayed protons would proceed not only to the g.s. in \isotope[63]{Ga} but also to one of the four excited states  marked in the left part of Fig. \ref{fig:as64_levsch}. Prior to our work, excited states in \isotope[63]{Ga} had  been studied in-beam  \cite{Weiszflog2001,rudolph_2021}. The levels observed here were previously reported  there except for the $925.1\mathrm{~keV}$ level which is also confirmed  in this work in the study of the  decay of \isotope[63]{Ge}, see below. The known levels and $\gamma$ transitions are shown in black in Fig.\ref{fig:as64_levsch} left, and the new level and its de-excitation in red. It is clear from the data that overall the proton decays favour the population of the $442.8\mathrm{~keV}$ and $75.0\mathrm{~keV}$ levels rather than the $722.2\mathrm{~keV}$ and $925.1\mathrm{~keV}$ levels, see Fig. \ref{fig:as64_pgated_gammaspectrum}. 
	Unfortunately the limited statistics did not allow the association of individual proton and $\gamma$ lines.
	The excitation energy region corresponding to the beta-delayed proton decay is marked as the shaded region in the \isotope[64]{Ge} level scheme.
	
	%%%%%%%%% fig 12 64As proton gated gamma spectrum %%%%%%%%%%%%%%%%%%%%

\begin{figure*}[!t]
		\includegraphics[width=12.2cm]{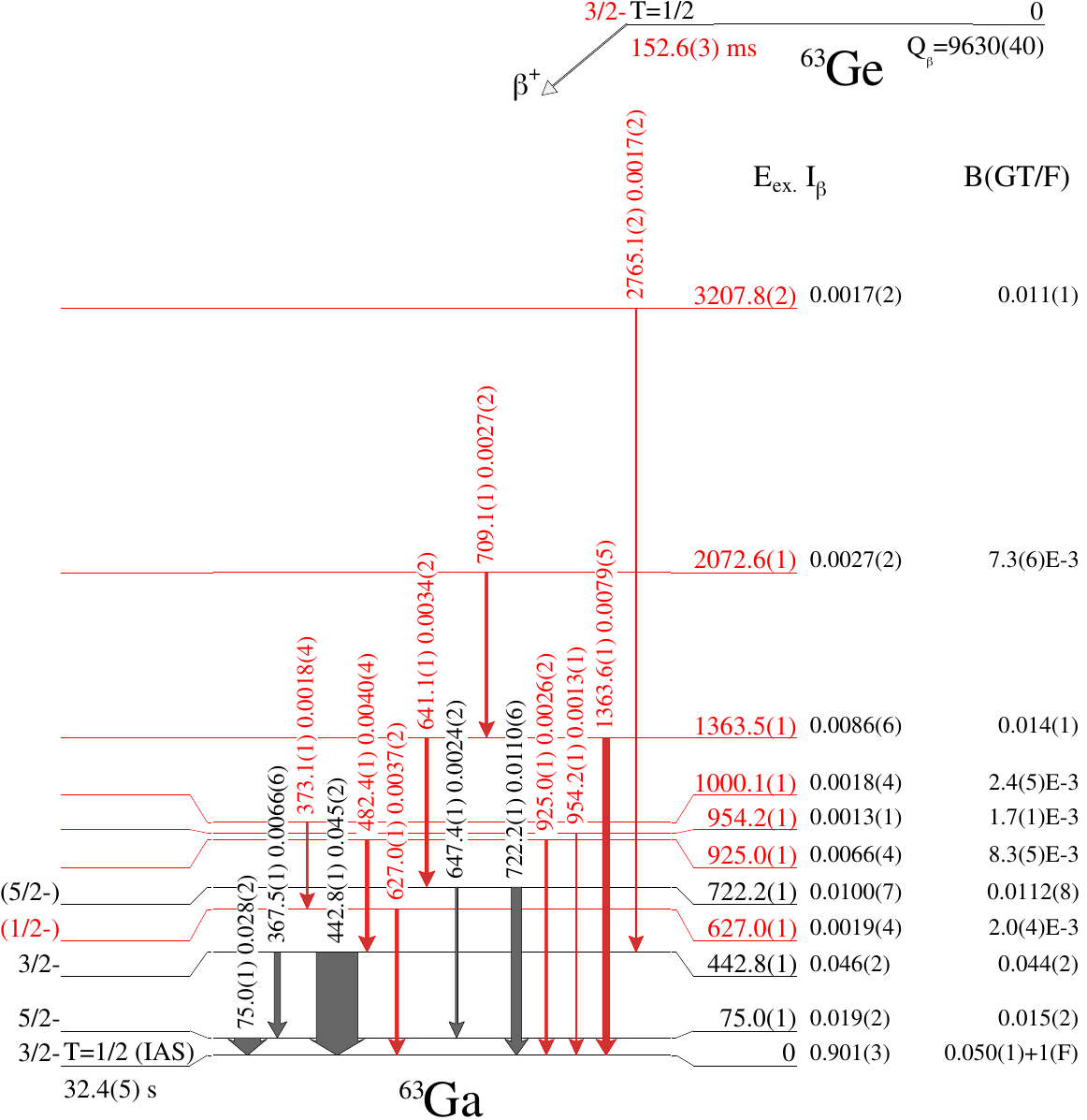}
		\caption{
			Decay scheme for \isotope[63]{Ge} based on the present experiment. Transition intensities are given per decay. B(GT/F) values are calculated  using our measured half-life and the $ Q_\beta $ value from the  literature \cite{wang_ame_2021}.
			Lines shown in red represent $\gamma$-rays and states observed for the first time. For levels without spin-parity assignment, allowed Gamow-Teller transitions are assumed for the B(GT) calculations. The Gamow-Teller contribution for the g.s.-g.s. beta decay was estimated assuming the expected value of 1 unit for the Fermi component, see text. Half-life of \isotope[63]{Ga} is from \cite{nubase2020}.
			\label{fig:ge63_levsch}
		}
	\end{figure*}

	The decay of \isotope[63]{Ge} was studied in a similar way to \isotope[64]{Se} and \isotope[64]{As}, namely by placing conditions for this nucleus on the ID plot and looking at the DSSSD or EURICA signals within a correlation time of $1050 \mathrm{~ms}$. 
	The resulting $\gamma$ spectrum is shown in Fig. \ref{fig:ge63_gamma_spectrum}.
	Three of the previously known levels were also observed here and are drawn in black in Fig. \ref{fig:ge63_levsch}. The $925.1\mathrm{~keV}$ level mentioned above was also seen in this decay.
	On the basis of $\gamma$-$\gamma$ coincidences we determined the existence of seven new excited states in \isotope[63]{Ga}, which are shown in red in the same figure. As in the \isotope[64]{As} case there is the possibility of populating levels that decay directly to the g.s. However the possible candidates were too weak in this case and we have not included them in the level scheme. The properties of the levels and transitions in \isotope[63]{Ga} are summarised in Table \ref{Tab:intensities_63Ge_decay}.

	%%%%%%%%% fig 13   63Ge antianalog decay  %%%%%%%%%%%%%%%%%%%%
	%\begin{turnpage}

	%\end{turnpage}
	
	%\begin{widetext}
	%\begin{turnpage}
	\begin{table*}
		\caption{ 
			The table shows the properties of the levels in \isotope[63]{Ga} populated in the $\beta$ decay of \isotope[63]{Ge} (see Fig. \ref{fig:ge63_levsch}). Column 1 indicates the energies of the levels in \isotope[63]{Ga}, column 2 the assigned spin-parity, columns 3 and 4 the $\gamma$-ray energy and intensity per decay,  column 5 the $\beta$ feeding to the level and column 6 the B(GT) or B(F) values. For levels without spin-parity assignment, allowed Gamow-Teller transitions are assumed for the B(GT) calculations. The Gamow-Teller contribution for the g.s.-g.s. beta decay was estimated assuming the expected value of 1 unit for the Fermi component, see text. The properties of the IAS are marked with numbers in bold.
			\label{Tab:intensities_63Ge_decay}}
		
		\begin{ruledtabular}
			\begin{tabular}{llllll}
				
				$E_{ex}$ keV&  $J_\pi$  &  $E_\gamma$ keV  &  $I_\gamma$   &  Feeding      &  $B(GT/F)$        \\\hline\hline
				\bf 0.0 IAS & $\mathbf{3/2^- ,\, T\!=\!+1/2}$ 
				& \bf           &  \bf          &  \bf 0.901(3) &  \bf 1(F) + 0.050(1)(GT)\\\hline
				75.0(1)     &  $5/2^-$   &  75.0(1)      &  0.028(2)     &  0.019(2)     &  0.015(2)          \\\hline               
				442.8(1)    &  $3/2^-$   &  442.8(1)     &  0.045(2)     &  0.046(2)     &  0.044(2)         \\
				&           &  367.5(1)     &  0.0066(6)    &               &           \\\hline
				627.0(1)    &  $(1/2^-)$ &  627.0(1)     &  0.0037(2)    &  0.0019(4)    &  0.0020(4)         \\\hline
				722.2(1)    &  $(5/2^-)$ &  647.4(1)     &  0.0024(2)    &  0.0100(7)    &  0.0112(8)         \\
				&           &  722.2(1)     &  0.0110(6)    &               &                   \\\hline
				925.0(1)    &           &  482.4(1)     &  0.0040(4)    &  0.0066(4)    &  0.0083(5)         \\
				&           &  925.0(1)     &  0.0026(2)    &               &                   \\\hline
				954.2(1)    &           &  954.2(1)     &  0.0013(1)    &  0.0013(1)    &  0.0017(1)        \\\hline
				1000.1(1)   &           &  373.1(1)     &  0.0018(4)    &  0.0018(4)    &  0.0024(5)         \\\hline
				1363.5(1)   &           &  641.1(1)     &  0.0034(2)    &  0.0086(6)    &  0.014(1)         \\
				&           &  1363.6(1)    &  0.0079(5)    &               &                   \\\hline
				2072.6(1)   &           &  709.1(1)     &  0.0027(2)    &  0.0027(2)    &  0.0073(6)         \\\hline
				3207.8(2)   &           &  2765.1(2)    &  0.0017(2)    &  0.0017(2)    &  0.011(1)         \\
			\end{tabular}
		\end{ruledtabular}
	\end{table*}
	%\end{turnpage}
	%\end{widetext}

	%****************************************************
	\subsection{Half-lives and proton branching ratios }
	%****************************************************
	\label{chap:T12 and Br}
	Half-life values are necessary to obtain the $ B(F) $ and $ B(GT) $ strengths of transitions to states populated in the daughter nucleus. 
	Moreover, they are of utmost importance in any kind of nuclear astrophysical process calculations. In particular, in  the present case they are relevant for p-process calculations \cite{zhou_mass_2023}. In the kind of experiment presented here,  the $T_{1/2}$ is determined by looking at the correlation time between the implantation of the ion  and a characteristic radiation emitted in the decay of the nucleus under study. If this radiation involves only the nucleus of interest one can use the simple decay formula to fit the data and extract the $T_{1/2}$.  
	However, if the radiation involves the nucleus under study and its descendants, the Bateman equations have to be used, and several parameters have to be taken into account. 
	In our analysis we have used both methodologies. 
	As can be seen in Fig. \ref{fig:decay_scheme_colors}, the decay radiation chain of \isotope[64]{Se} involves descendants of the $A\!=\!64$ and the $A\!=\!63$ chains. 
	However, from all of them, only \isotope[64]{Se}, \isotope[64]{As} and \isotope[63]{Ge} have sub-second half-lives and are relevant to the present discussion. 
	We will first discuss  the \isotope[64]{As} and \isotope[63]{Ge} cases, which are simpler and a necessary step before we discuss the \isotope[64]{Se} $T_{1/2}$ and proton branching.

	%\clearpage
	%****************************************************
	\subsubsection{Half-life and proton branching of \isotope[64]{As}}
\begin{figure}
		\includegraphics[width=8.6cm]{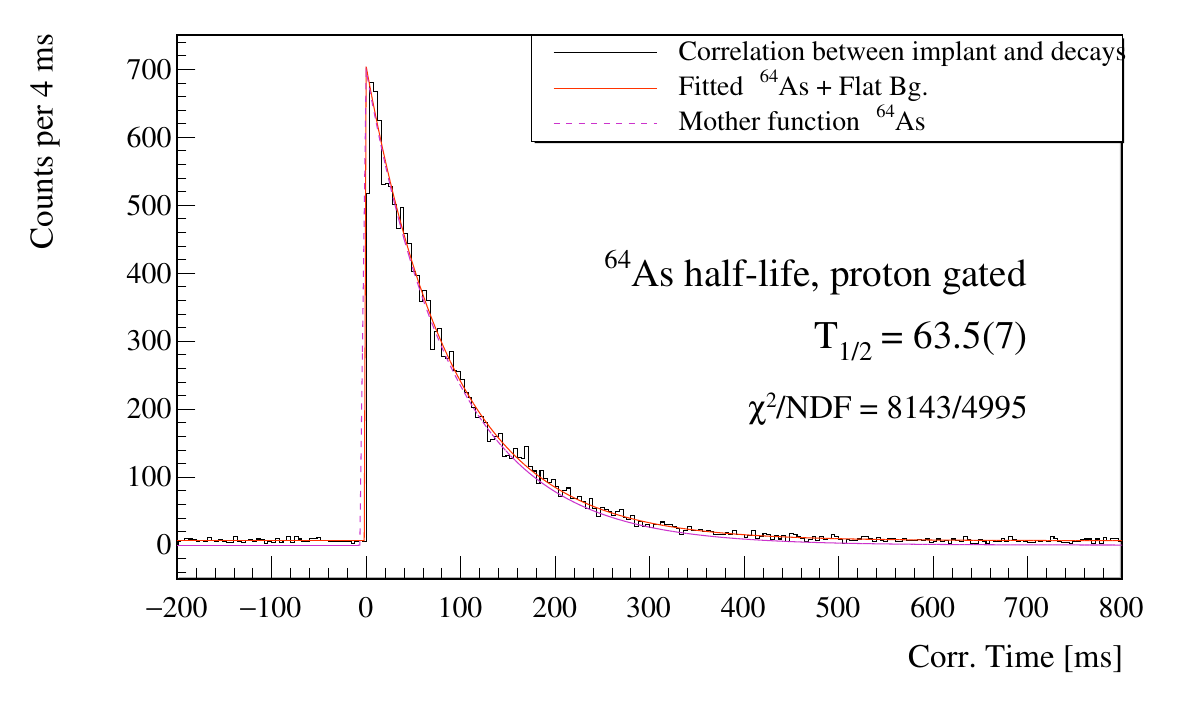}
		\caption{
			Correlation spectrum between  \isotope[64]{As} implants and detected protons in the same pixel (see text). The red solid line shows a fit to an exponential function and a flat background. 
			\label{fig:as64_T12_fit}
		}
	\end{figure}

	%****************************************************
	As can be seen in the DSSSD spectrum with the condition on \isotope[64]{As} implantation shown in Fig. \ref{fig:as64_pspectrum}, there is a small proton branching following this decay. 
	The protons can be recognised as the typical structure peaks above 1.2 MeV. 
	The first value for the $T_{1/2}$ of \isotope[64]{As} was obtained from the time correlation of the \isotope[64]{As} implantation with the DSSSD decay signal above $1254\mathrm{~keV}$. 
	This could  be fitted with a simple decay formula and the value obtained was $63.5(7)\mathrm{~ms}$ (see Fig. \ref{fig:as64_T12_fit} and also Table \ref{Tab:halflives}).

	%%%%%%%%% fig 14    64As  T12  %%%%%%%%%%%%%%%%%%%%

	%\begin{turnpage}
	\begin{table}
		\caption{ The table shows a summary of results for the three isotopes in the \isotope[64]{Se} decay chain. Column 1 shows the number of implantations, columns  2, 3 and 4 show the $T_{1/2}$ values obtained using the various methods of analysis as described in section \ref{chap:T12 and Br}.
			We have adopted the $T_{1/2}^\mathrm{DSSSD}$ value as the final half-life value in our analysis, since the three values shown here are not fully independent, 
			because they are based on partially overlapping sets of data. These values are shown in bold.
			The  proton branching ratios are shown in the last column.\label{Tab:halflives}}
		
		\begin{ruledtabular}
			\begin{tabular}{cp{8ex}ccp{10ex}}
				\multicolumn{5}{c}{\bf \isotope[63]{Ge}}   \\
				Imp. Ions & &   $T_{1/2}^{\gamma}$   &  $\boldsymbol{T_{1/2}^\textbf{DSSSD}}$ &   \\
				3.860.972& & $152.7(28)$   &  $\boldsymbol{152.6(3)}$     &      
			\end{tabular}
			\begin{tabular}{ccccc}
				\multicolumn{5}{c}{\bf \isotope[64]{As}}   \\
				Imp. Ions &$T_{1/2}^\mathrm{p}$ &   $T_{1/2}^{\gamma}$  &  $\boldsymbol{T_{1/2}^\mathbf{DSSSD}}$   & $B_{p}$  \\
				481.896 & $63.5(7)$    & $62.7(7)$    &   $\boldsymbol{63.0(3)}$     & $4.4(1)\%$ 
			\end{tabular}
			\begin{tabular}{ccccc}
				\multicolumn{5}{c}{\bf \isotope[64]{Se}}   \\
				Imp. Ions &$T_{1/2}^\mathrm{p}$ &   $T_{1/2}^{\gamma}$   &  $\boldsymbol{T_{1/2}^\mathbf{DSSSD}}$  & $B_{p}$  \\
				10.308&$23.3(6)$     & $20.0(31)$    &   $\boldsymbol{22.5(6)}$     & $48.0(9)\%$
			\end{tabular}
		\end{ruledtabular}
	\end{table}
	%\end{turnpage}
	
\begin{figure}
		\includegraphics[width=8.6cm]{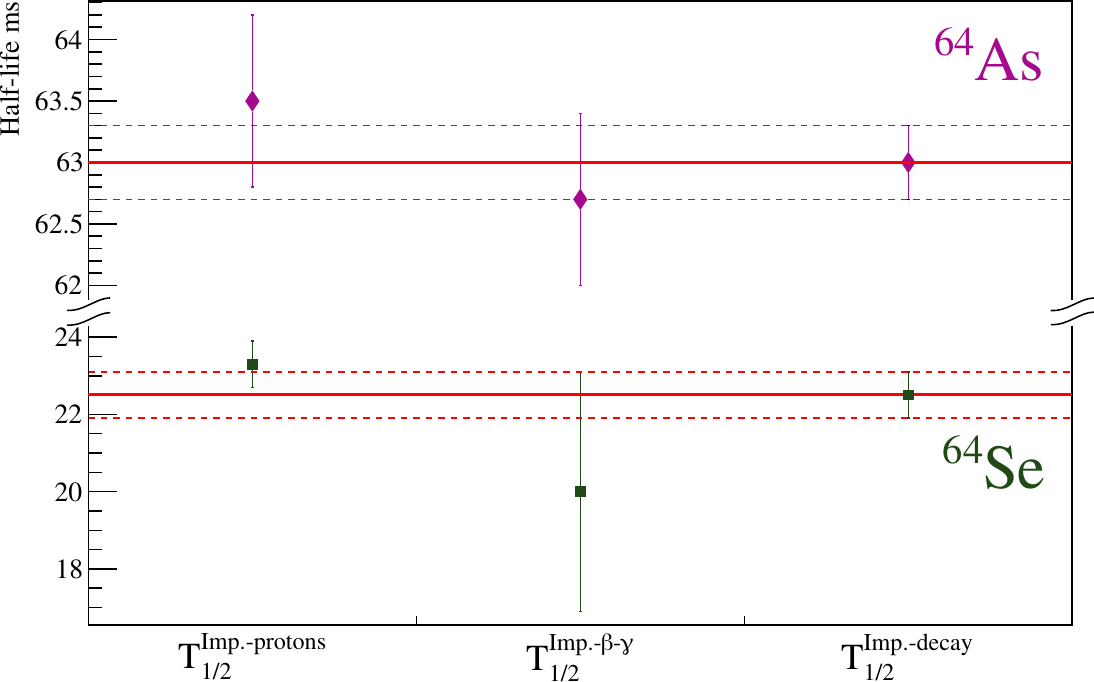}
		\caption{
			Measured values of the half-lives of  \isotope[64]{As} and \isotope[64]{Se}, marked with  $\blacklozenge$ and $\blacksquare$, obtained using implantation-beta-proton,  implantation-beta-gamma correlations and implantation-decay, as described in the text. The solid lines mark the weighted average of the $T_{1/2}$ values of the two isotopes with the corresponding  errors shown as  dashed lines. 
			\label{fig:se64_as64_halflives}
		}
	\end{figure}

	From this fit one can also obtain the total number of \isotope[64]{As} ions that decay by protons, which, after dead time correction and  comparison with the number of \isotope[64]{As} implants provided a proton branching ratio of $B_{p}(\isotope[64]{As})=4.4(1)\%$, also included in Table \ref{Tab:halflives}. 
	Knowing this number one can now correlate all the DSSSD signals, above thresholds of $55\mathrm{~keV}$ (X-side) and $133\mathrm{~keV}$ (Y-side), in other words including $\beta$ particles and protons, and obtain a second value for the $T_{1/2}$ of \isotope[64]{As} of $63.0(3)\mathrm{~ms}$ and a $\beta$ efficiency of $\varepsilon_{\beta}$ of $83.6(6)\%$. 
	The latter was obtained by comparing the total number of implants with the detected $\beta$ particles, taking into consideration the
	proton branching.
	Finally, one can correlate the implants with the characteristic $\gamma$ radiation of \isotope[64]{As}
	decay in \isotope[64]{Ge}, namely the $676.9\mathrm{~keV}$, $901.4\mathrm{~keV}$, $1495.1\mathrm{~keV}$ and $1888.4\mathrm{~keV}$ $\gamma$ rays, obtaining thus a $T_{1/2}$ of $62.7(7)\mathrm{~ms}$.
	All of these values are summarised in Table \ref{Tab:halflives} and plotted in Fig. \ref{fig:se64_as64_halflives}. 
	We have taken for our calculations the $T_{1/2}^\textrm{DSSSD}$ value since it has the smallest error from all three.
	Combining these values is difficult because they share, at least partially, the same data set.
	They agree within two standard deviations  with previously reported values of $72(6) \mathrm{~ms}$ \cite{rogers2014} and $18^{+43}_{-7}\mathrm{~ms}$  \cite{LopezJimenez2002}, and differ slightly from a value of $ 69.0(14) \mathrm{~ms} $  \cite{giovinazzo2020} based on an early analysis of part of the data taken in this campaign.
	
	%%%%%%%%% fig 17    64Se and 64As T12  %%%%%%%%%%%%%%%%%%%%

	%****************************************************
	\subsubsection{Half-life  of \isotope[63]{Ge}}
	%****************************************************
	
\begin{figure}
		\includegraphics[width=8.6cm]{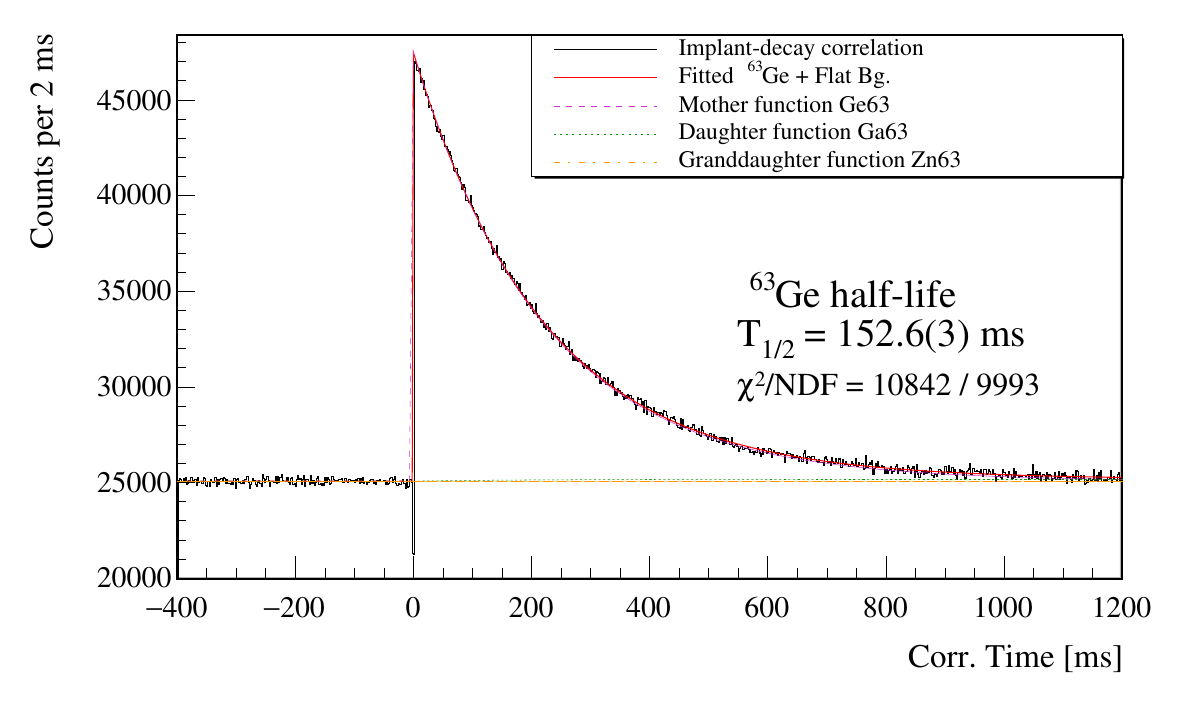}
		\caption{Correlation spectrum between  \isotope[63]{Ge} implants and betas detected in the same pixel (see text). The fit shown as a solid red line includes the mother and daughter activities  and a flat background.
			\label{fig:ge63_T12_fit}
		}
	\end{figure}

	The $\beta$ decay of \isotope[63]{Ge} proceeds to the \isotope[63]{Ga} ground state or to excited states in this nucleus with the subsequent $\beta$ - delayed $\gamma$ radiation. 
	In this case we obtained a $T_{1/2}$ value of  $152.6(3) \mathrm{~ms}$ and an $\varepsilon_{\beta}$  value of $82.7(2)\%$ from the implantation-$\beta$ correlations (see Fig. \ref{fig:ge63_T12_fit}),  and a $T_{1/2}$ value of $152.7(28) \mathrm{~ms}$ from the correlation of the implantation events with the $442.8\mathrm{~keV}$ $\gamma$ rays. 
	See Table \ref{Tab:halflives}. Our results are in reasonable agreement with the previous $150(9)\mathrm{~ms}$, $149(4)\mathrm{~ms}$ and $156(11)\mathrm{~ms}$ half-life values for \isotope[63]{Ge} based on two independent measurements performed at GANIL \cite{blank2002,rogers2014,kucuk2017}.
	
	%%%%%%%%% fig 15    63Ge  T12  %%%%%%%%%%%%%%%%%%%%
	
	%****************************************************
	\subsubsection{Half-life and proton branching of \isotope[64]{Se}}
	%****************************************************
	
\begin{figure}
		\includegraphics[width=8.6cm]{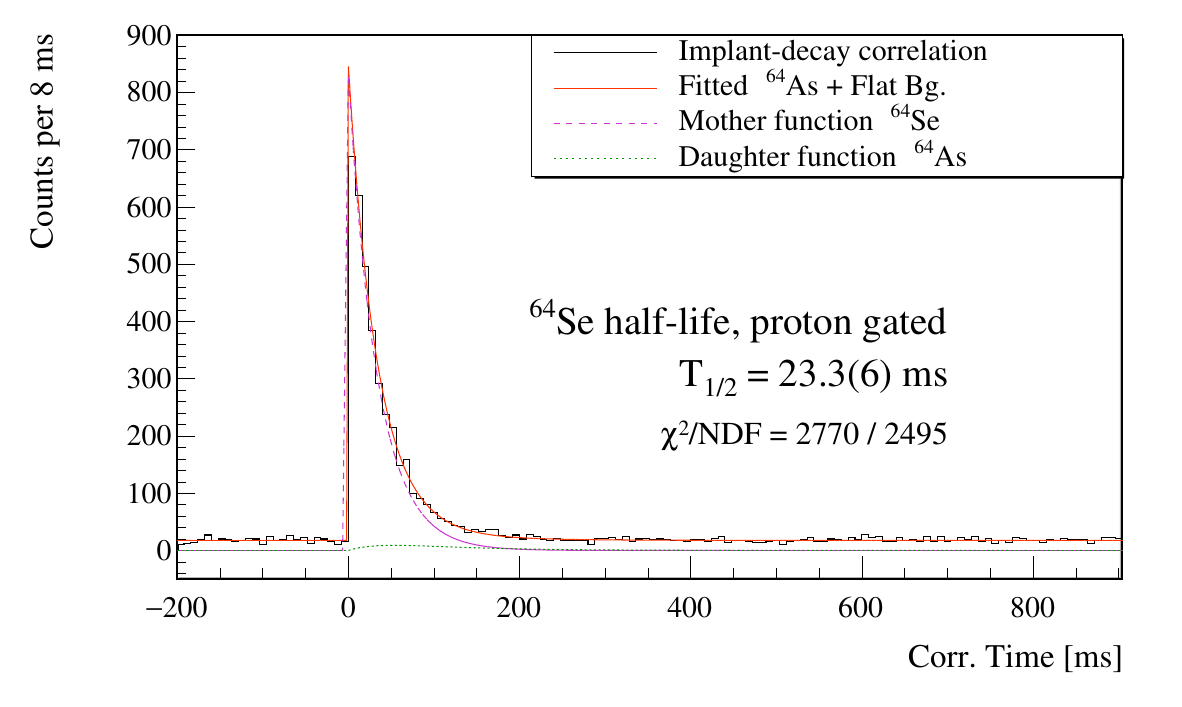}
		\caption{
			Correlation spectrum between  \isotope[64]{Se} implants and detected protons in the same pixel (see text). The red solid line represents a fit that includes the mother and daugther activities  and a flat background.  
			\label{fig:Se64_T12_fit}
		}
	\end{figure}

	Finally, we have analysed the \isotope[64]{Se} decay. Prior to this work, the only experimental information on the half-life value was a lower limit of $ 180\mathrm{~ns} $ obtained from the time of flight of at least one event reported in Ref.  \cite{Stolz2005}.
	Looking in our experiment at the correlations between the implantation events and the proton signals detected in the DSSSD above $1420\mathrm{~keV}$ (see Fig. \ref{fig:Se64_T12_fit}), and taking into consideration the contribution from \isotope[64]{As} with the $T_{1/2}$ and $B_{p}$ obtained, we deduced a value of $23.3(6)\mathrm{~ms}$ for the $T_{1/2}$ of \isotope[64]{Se}, and a $B_{p}$ value of $48.0(9)\mathrm{~\%}$. 
	The latter value was obtained by comparing the number of \isotope[64]{Se} proton decays with the number of implanted ions for this isotope.
	One can also correlate the implants with the characteristic $\gamma$ radiation of \isotope[64]{Se}
	decay in \isotope[64]{As}, namely the $697.4(1)\mathrm{~keV}$, $1259.2(2)\mathrm{~keV}$ and $1449.2(2)\mathrm{~keV}$ $\gamma$ rays, obtaining thus a $T_{1/2}$ of $20.0(31)\mathrm{~ms}$.
	Finally, analysing the full DSSSD spectrum, and including in the Bateman equations the values previously obtained for the $T_{1/2}$ and $B_{p}$ of \isotope[63]{Ge} and \isotope[64]{As}, $22.5(6) \mathrm{~ms}$ was obtained for the $T_{1/2}$ of \isotope[64]{Se} and a value for the $\varepsilon_{\beta}$ of $86.7(15)\mathrm{~\%}$. 
	
	All the $T_{1/2}$ values and the proton branching ratios are summarised in Table \ref{Tab:halflives} and the $T_{1/2}$ values are shown in Fig. \ref{fig:se64_as64_halflives}. To determine $\varepsilon_\beta$, we used the values explained above plus information obtained from other nuclei in the same experiment obtaining an average  value of $0.82(4)$, where the error corresponds to the standard deviation.
	
	%%%%%%%%% fig 16    64Se  T12  %%%%%%%%%%%%%%%%%%%%

	%****************************************************
	\section{Discussion of \isotope[\textbf{64}]{\textbf{S\lowercase{e}}} and \isotope[\textbf{64}]{\textbf{A\lowercase{s}}} ground state masses}
	%****************************************************
	\label{sec:masses}
	%****************************************************
	
	Before this experiment, the masses  of \isotope[64]{Se} and  \isotope[64]{As} were not known experimentally and the tabulated values were extrapolated values. As mentioned earlier, very recently a direct mass measurement has been carried out at Lanzhou \cite{zhou_mass_2023} providing the first experimental value for the
	\isotope[64]{As} mass, while the mass of \isotope[64]{Se} remains experimentally unreachable.
	As we have explained in Section \ref{sec:levelsch_64se},  we can connect the experimentally known mass of \isotope[63]{Ge} with the mass of 
	\isotope[64]{As} using the proton energy and the $\gamma$ information, see Fig. \ref{fig:MassScheme_63Ge-64As}. 
	Using this information we obtained a mass value of 
	\begin{equation}
		M^\mathrm{EXP}\left(\isotope[64]{As},\mathrm{g.s.} \right) = -39588(50)\mathrm{~keV}.
	\end{equation}
	This agrees with the extrapolated value  of $-39530\hashtag(200)\mathrm{~keV}$ from AME2020 \cite{wang_ame_2021} and also with the
	recently measured value of $-39710 (110)\mathrm{~keV}$ \cite{zhou_mass_2023}.
	
	From the point of view of the $ A\!=\!64 $ isobaric quintuplet ($ T\!=\!2 $) there are now three members with known masses, the ground state of \isotope[64]{Zn} and its  IAS state in  \isotope[64]{Ga}, and the IAS of \isotope[64]{Se}
	into \isotope[64]{As}. 
	Using these three mass excess values we obtain the $T\!=\!2$ Isobaric Mass Multiplet Equation (IMME),
	\begin{equation}
		M\left( T_z \right) = -47473(24) - 9644(35)~T_z + 189(12)~T_z^2.
	\end{equation}
	Evaluating this curve for $ T_z\!=\!-2 $ we obtain an estimated value of,
	\begin{equation}
		M^\mathrm{IMME}\left(\isotope[64]{Se},\mathrm{g.s.} \right) = -27429(88)\mathrm{~keV}
	\end{equation} 
	for the \isotope[64]{Se} mass excess. This value is to be compared with the AME2020 evaluated number, $-26860\hashtag(500)\mathrm{~keV}$, which suggests a defect in the estimation of the  binding energy in AME2020. We have observed a similar tendency  in other cases in this region. A summary of such a comparison can be found in Fig. 2 of Ref. \cite{Orrigo23}.

	%****************************************************
	\section{Discussions of decay schemes}
	%****************************************************
	\label{sec:discussions}
	%****************************************************
	
	In order to discuss the decay schemes of the nuclei we have studied, it is very important to know the Fermi (F) or Gamow-Teller (GT) strengths associated with the $\beta$ feeding to the individual levels.  They can be calculated using the following formulae, 
	
	\begin{align}
		B(F)  &= \frac{K~I_\beta}{f(Q_\beta-E,Z)~T_{1/2}}, \label{eq: B(F)}\\
		B(GT) &= \frac{K~I_\beta}{\lambda^2 ~f(Q_\beta-E,Z)~T_{1/2}} \label{eq: B(GT)}
	\end{align}
	where $ K\!=\!6144.5 (37) $ \cite{Hardy2020} and $ \lambda\!=\!-1.2761^{+14}_{-17} $ \cite{MUND2013,Abele2002}. The Fermi function $f$ can be obtained using the NNDC Logft tool \cite{nndc_logft}. The feedings are obtained from the $\gamma$ and proton intensities measured in this experiment.
	We have also used the $T_{1/2}$ values determined in our measurements and the $ Q_\beta $ values from the literature \cite{wang_ame_2021} together with our measured value of $ -39598(50)\mathrm{~keV} $ for the \isotope[64]{As} ground state and the deduced value of $ -27429(88)\mathrm{~keV} $ for  \isotope[64]{Se}.

	%****************************************************
	\subsection{ Decay of \isotope[\textbf{64}]{\textbf{Se}} } 
	%****************************************************
	\label{sec:analysis_64se}
	%****************************************************
	
	The \isotope[64]{Se} $\beta$ decay scheme is shown in Fig. \ref{fig:se64_levsch}. On the right hand side of the levels we show the direct feeding and associated $\beta$ strength  obtained with equations \ref{eq: B(F)} and \ref{eq: B(GT)}. 
	As discussed in section \ref{sec:levelsch_64se}, we have identified the IAS as the level at $1956.5\mathrm{~keV}$ excitation energy, which decays by proton emission as well as by electromagnetic radiation. This is because the proton decay is isospin forbidden and consequently severely hindered. The $B(F)$ for the transition to this level is $3.8 (3)$, very close to the expected value of 4. The reader should note that we were not able to observe conversion electrons in this experiment. 
	Consequently, a possible E0 transition from the $0^+$ IAS to the $0^+$ ground state is not taken into account, but the $B(F)$ value for the $1956.5\mathrm{~keV}$ level is likely to be closer to the expected value of 4. The ground state of \isotope[64]{As} has been assigned $0^{+}$ based on isospin symmetry and the comparison with \isotope[64]{Ga}. 
	As we shall see later in this section, our data corroborate the good symmetry between these two nuclei at low excitation energy which reaffirms the $0^{+}$ assignment. Moreover, the proposed decay scheme, and in particular the de-excitation of the populated levels, is in accord with this assignment.
	In principle, a very small admixture of the T=2 IAS could be present in the ground state of \isotope[64]{As}, as has been observed in the mirror nucleus \isotope[64]{Ga}, \cite{RAMAN1975} but a similar mixing in our case would translate to a very small feeding to the ground state,   well below our sensitivity. The apparent feeding we observe to the ground state,  is probably due to some unobserved feeding to states at higher energy that de-excite to the ground state, or to the above mentioned unobserved E0 transitions.

\begin{figure}
		\includegraphics[width=8.6cm]{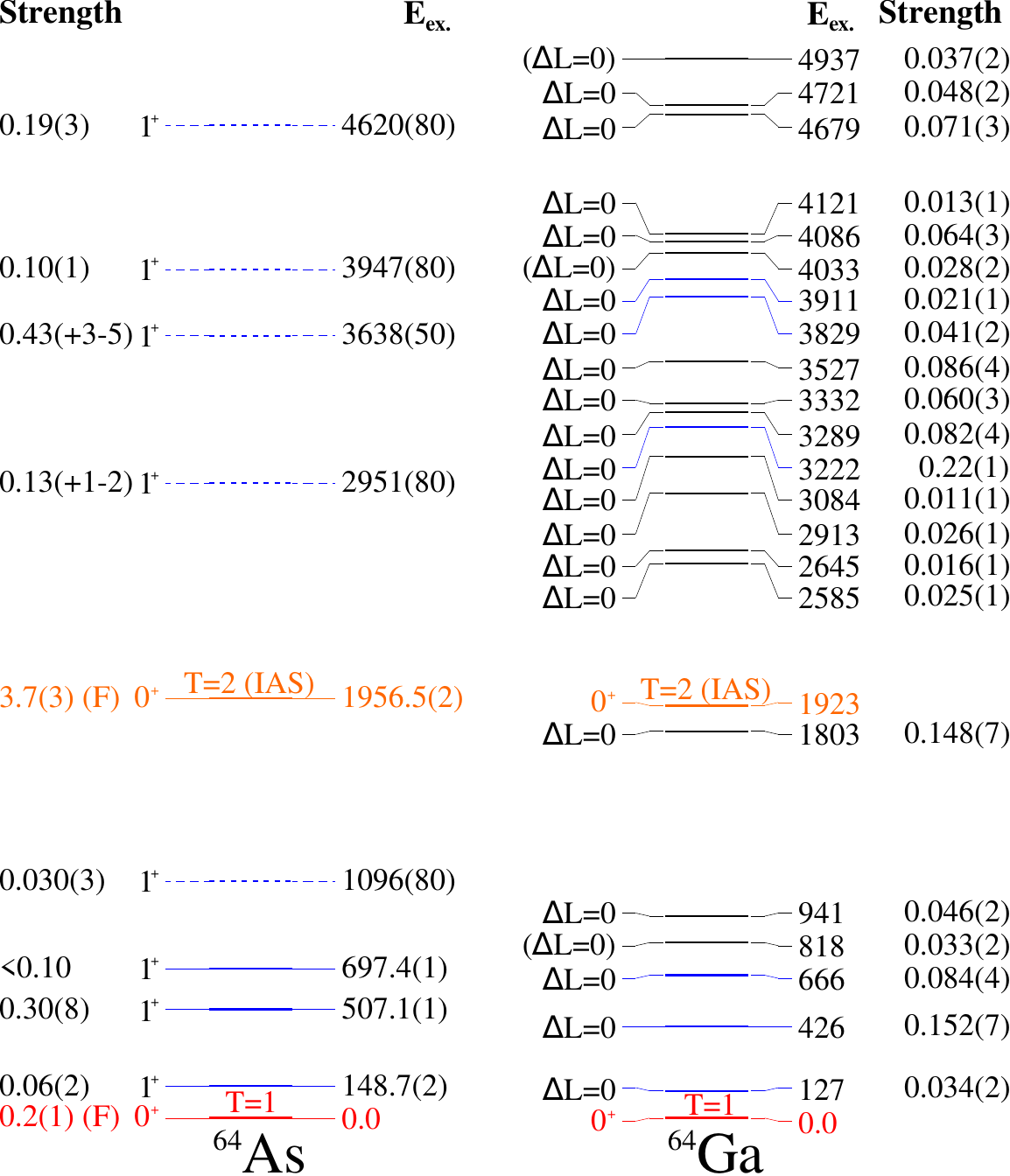}
		\caption{
			A comparison of the  levels in \isotope[64]{As} and \isotope[64]{Ga}, fed respectively by  beta decay and the charge exchange $ \left(\isotope[3]{He},t\right) $ reaction \cite{Diel2015,Diel2019}. The strengths observed in both studies are also shown. The states in
			\isotope[64]{Ga} are shown on the right and those that we think correspond with states in the mirror nucleus are shown in blue.
			\label{fig:CEcomparison_levsch}
		}
	\end{figure}

	No other levels with $0^{+}$ spin-parity are expected to be directly populated in the decay. 
	In consequence, all levels with direct feeding are tentatively assigned as $1^{+}$, except for the 148.7, 507.1, and 697.4 keV levels, where a clear correspondence with levels at similar energies are found in the mirror nucleus \isotope[64]{Ga},  and the $1^{+}$ assignment is considered firm (see the discussion below and Fig. \ref{fig:CEcomparison_levsch}).

	In the beta-delayed proton emission of \isotope[64]{Se}, we populate one excited state in \isotope[63]{Ge} de-exciting by a $417.5\mathrm{~keV}$ $\gamma$ transition.
	This energy is very similar to the $442\mathrm{~keV}$ transition  in \isotope[63]{Ga}, and is therefore assumed to be the mirror transition in this discussion. The $442.8\mathrm{~keV}$ transition in \isotope[63]{Ga} has been observed in \cite{Weiszflog2001,rudolph_2021} where it was assigned spin and parity ($3/2^-$). In our experiment, the $3/2^-$ ground state (see section \ref{sec:analysis_63ge} for spin-parity assignment to the \isotope[63]{Ge} ground state) and the state under discussion, at $417.5\mathrm{~keV}$ are populated in the proton-decay of the IAS. 
	The intensity to the ground state is double the intensity to the $417.5 \mathrm{~keV}$ state. Since the population of the ground state requires an $L\!=\!1$ angular momentum transfer, the most probable scenario is that the proton decay to the $417.5 \mathrm{~keV}$ excited state also involves an $L\!=\!1$ transfer. Taking into consideration that this level is fed from a $7/2^-$ state in the mirror \cite{Weiszflog2001}, favours the spin-parity of the $417.5 \mathrm{~keV}$ state, and consequently of the $442.8 \mathrm{~keV}$ state, being $3/2^-$, as suggested in \cite{rudolph_2021}.
	We note here that the present experiment provides the only possibility of populating excited states in \isotope[63]{Ge} via beta-decay since the hypothetical parent nucleus \isotope[63]{As}, probably does not exist as a bound system. Nevertheless $\gamma$ radiation of $417.5 \mathrm{~keV}$ energy has been observed 
	\cite{henry_thesis_2015} in a two neutron knock out reaction on \isotope[65]{Ge} leading to \isotope[63]{Ge}, thus reinforcing our identification.

\begin{figure}
		\includegraphics[width=8.6cm]{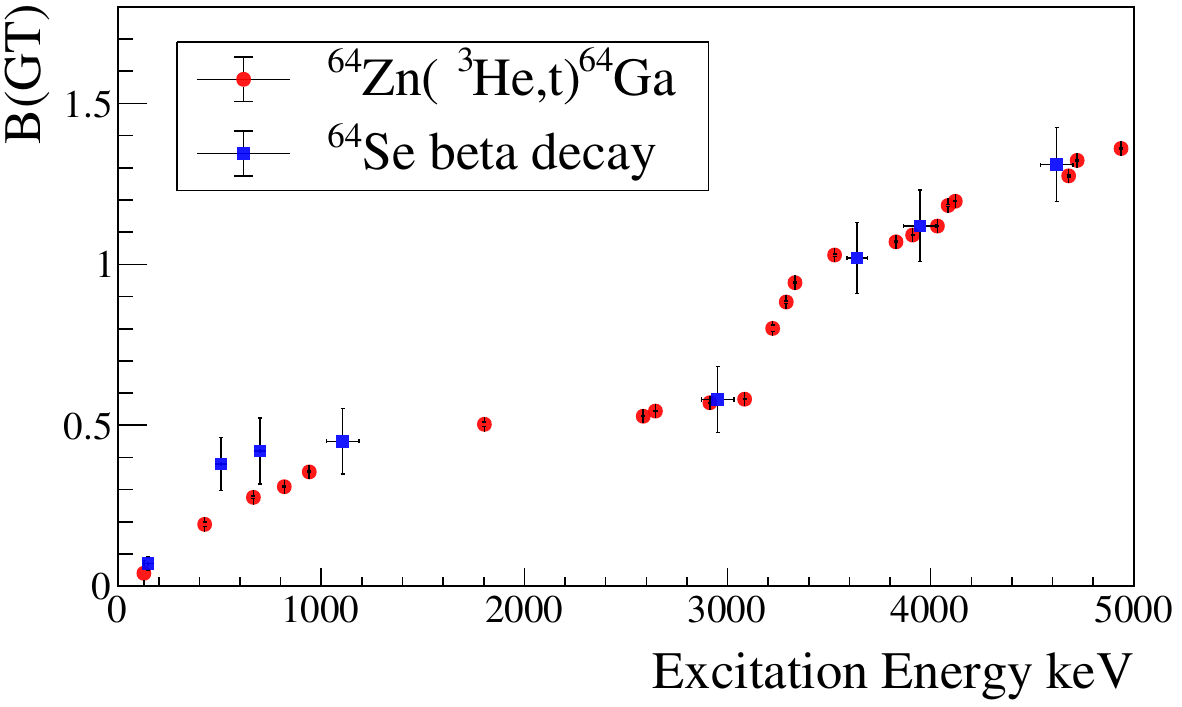}
		\caption{
			Comparison of the 
			measured cumulative B(GT) values in \isotope[64]{Se} beta decay  and the corresponding charge exchange reaction on the mirror nucleus \isotope[64]{Zn} \cite{Diel2015,Diel2019}. \label{fig:BGT_cumulative}   }
	\end{figure}

	As mentioned in the introduction, \isotope[64]{Se} is the heaviest $ T_z\!=\!-2 $ decay where we can readily make the comparison with the $ T_z\!=\!+2 $ mirror process $ \isotope[64]{Zn}\!\left(\isotope[3]{He},t\right)\!\isotope[64]{Ga} $. 
	The measurement was carried out at RCNP (Osaka, Japan) and published in \cite{Diel2015,Diel2019}. 
	The comparison between the levels populated in \isotope[64]{Se} by $\beta$ decay and \isotope[64]{Zn} in the charge exchange reaction, together with the $B(GT)$ is shown in Fig. \ref{fig:CEcomparison_levsch}. 
	Relatively good isospin symmetry is observed up to and including the excitation energy of the IAS in both nuclei. 
	Two more levels were observed in CE with respect to the beta decay in this region of energy. 
	In the higher part of the CE spectrum it is clear that the excellent resolution of $40\mathrm{~keV}$  achieved at Osaka, allows one to resolve more levels than in the beta decay. Regarding the B(GT) strength, differences are observed if one looks at individual levels. However, excellent agreement is found if we look at the cumulative strength (see Fig. \ref{fig:BGT_cumulative}).
	This is probably an indication of the existence of more proton peak contributions than the small number used in the fit.

	%%%%%%%%% fig 20    64Se and 64Zn mirror comparison %%%%%%%%%%%%%%%%%%%%

	%%%%%%%%% fig 19 cumulative bgt %%%%%%%%%%%%%%%%%%%%
	%\begin{turnpage}
	%\end{turnpage}

	%****************************************************
	\subsection{ Decay of \isotope[\textbf{64}]{\textbf{As}} }
	%****************************************************
	\label{sec:analysis_64as}
	%****************************************************
	
	The decay of \isotope[64]{As} is a very interesting case because its daughter nucleus, \isotope[64]{Ge}, is an $N\!=\!Z$ nucleus. 
	In fact, this nucleus has attracted considerable attention in the past in the search for Isospin forbidden E1 transitions \cite{Ennis1991,Farnea2003}, which were finally observed  by Farnea et al.  in \cite{Farnea2003}.
	In our experiment, \isotope[64]{As} was produced with considerable intensity and therefore a detailed decay scheme could be constructed. 
	This is presented in Fig. \ref{fig:as64_levsch} where the previously known levels from in-beam studies are marked in black \cite{Ennis1991,Farnea2003}.
	As is typical of decay data we populate levels with a different range of $\mathrm{J}^{\pi}$ and level energies than observed in in-beam studies, so the overlap is only at low excitation energy.
	In this experiment we have extended considerably the knowledge of excited levels of low spin up to $5\mathrm{~MeV}$, based on the observed $\gamma$-rays.
	We have also observed a small fraction, 4.4(1)$\%$, of $\beta$-delayed protons. They are shown in Fig. \ref{fig:as64_pspectrum} and marked in the \isotope[64]{Ge} scheme, Fig. \ref{fig:as64_levsch}, as a shaded area.

	As usual, we sought the most intensely populated state in this decay, which turns out to be the $4962.6(2)\mathrm{~keV}$ state. Accordingly we believe this state to be the IAS with spin and parity $0^{+}$  and $T\!=\!1$. It decays predominantly to an intermediate $1^{+}$ state at $3466.9(1)\mathrm{~keV}$ excitation energy which decays, in turn, to a known yrast and a near yrast $2^{+}$ state. We observe  that the IAS is populated with an intensity that corresponds to a B(F) value of $1.51(9)$. This is less than the expected value of 2 units. Similar to the \isotope[64]{Se} case, an E0 transition to the ground state could exist, and we would not have been able to detect it. 
	
	Among the other observed states one is clearly strongly populated, the level at $3466.9(1)\mathrm{~keV}$ with 0.12 apparent feeding, and we have assigned it as $1^+$. 
	The level at $2154.6(2)\mathrm{~keV}$ was also observed by Ennis et al. \cite{Ennis1991}, and assigned ($4^+$) spin-parity. In our experiment this level is fed by the above discussed $1^+$ state (see Fig. \ref{fig:as64_levsch}) making the $(4^+)$ assignment unlikely. Considering that it decays to the $2_2^+$ state at $1578.4(1)\mathrm{~keV}$  and not to the $0^+$ ground state,  we propose $(3^+)$.
	The existence of a $3^+$ state at this energy is possible given that this nucleus is doubly magic in triaxial space. See Fig. 46 in \cite{RAGNARSSON19781} and corresponding discussion. A possible $0^+$ assignment to the level at 2955.8 keV  will be discussed in section \ref{sec:multiplet discussion}.
	
	A number of other states are weakly populated and we have not assigned spin parities. 
	
	\section{Discusion of the T=2 isobaric multiplet}
	\label{sec:multiplet discussion}
	
\begin{figure*}
		\includegraphics[width=17.2cm]{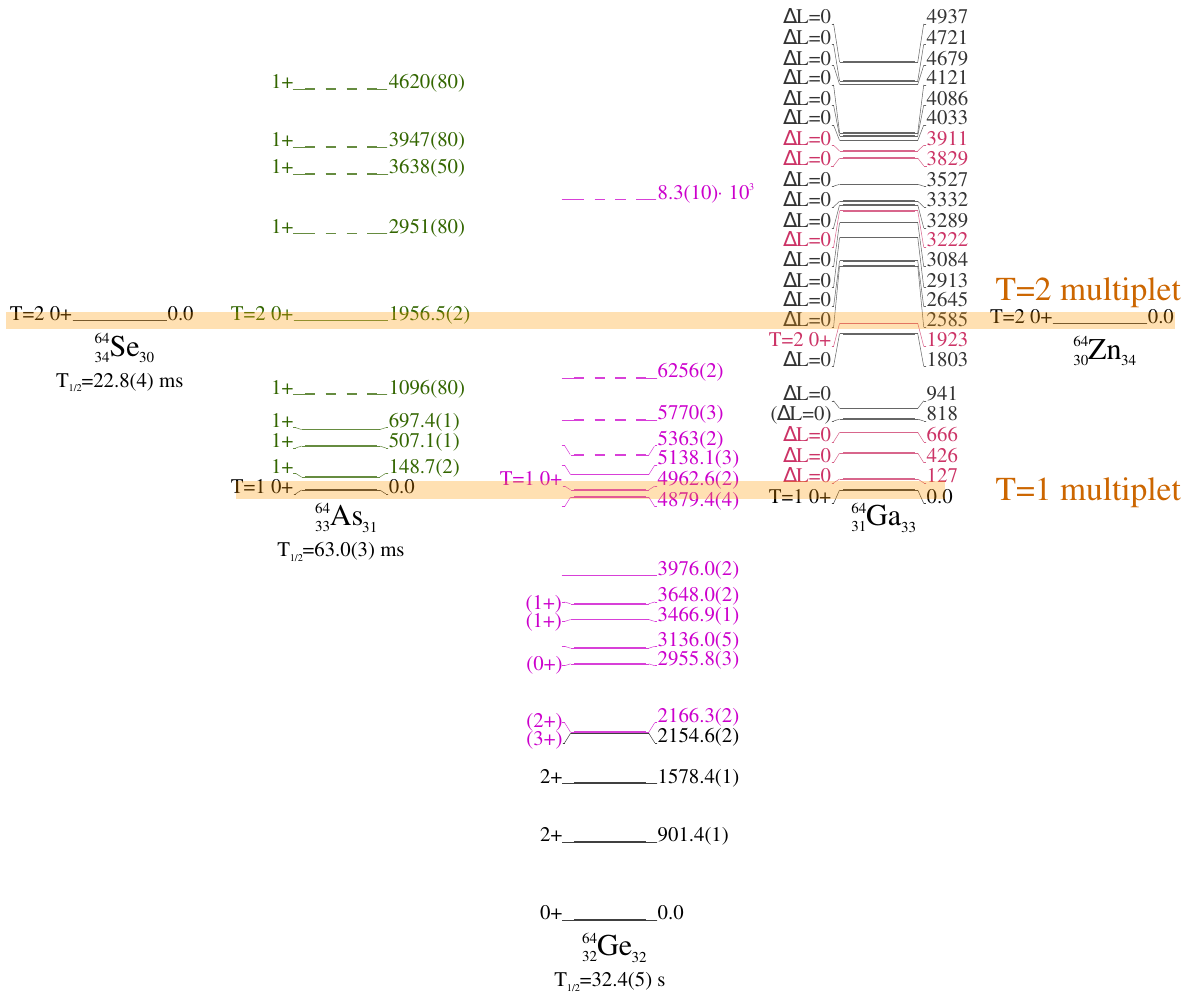}
		\caption{
			The T=2 isobaric multiplet
			overview. The figure shows a schematic view of the relationship of levels in the $T_z\!=\!-2$ to $T_z\!=\!+2$ $A\!=\!64$ nuclei.  Coulomb energy differences have been removed and so the four $T\!=\!2$  $0^+$ IASs appear at the same level. The two  $T_z\!=\!-1$ and $T_z\!=\!+1$ members of the T=1 $0^+$ multiplet also appear to be aligned, while the $T\!=\!1$ $0^+$ member in the $T_z\!=\!0$ nucleus 
			\isotope[64]{Ge} has been aligned with the other T=1 IAS states. See text. Both the $T\!=\!2$
			and $T\!=\!1$ multiplets are highlighted in orange.
			\label{fig:beautiful_multiplet_figure}
		}
	\end{figure*}

	A summary of our observations on the \isotope[64]{Se} $\rightarrow$ \isotope[64]{Ar} $\rightarrow$ \isotope[64]{Ge} decay chain, together with the information from the $ \isotope[64]{Zn}\!\left(\isotope[3]{He},t\right)\!\isotope[64]{Ga} $ experiment
	performed at Osaka \cite{Diel2019}, is presented in Fig. \ref{fig:beautiful_multiplet_figure}.

	%%%%%%%%% fig 21   beautiful multiplet %%%%%%%%%%%%%%%%%%%%

	In Fig. \ref{fig:beautiful_multiplet_figure} we have aligned the four known members of the $T\!=\!2$ multiplet. Because of the excellent symmetry observed for the two IAS states in \isotope[64]{As} and \isotope[64]{Ga}, the two extreme members of the $T\!=\!1$ multiplet, namely the ground states of \isotope[64]{As} and \isotope[64]{Ga} also appear to be naturally aligned.
	We have also included our results on the decay of \isotope[64]As into \isotope[64]{Ge}, aligning the IAS of the $T\!=\!1$ \isotope[64]{As} in \isotope[64]{Ge} with the ground state of \isotope[64]{As}, as shown in Fig. \ref{fig:beautiful_multiplet_figure}. 
	As depicted in the figure, this work has substantially contributed to our understanding of the $T\!=\!2$ and $T\!=\!1$ multiplets. It should be noted that this is the last case where such a comprehensive comparison can be readily made today  because the heavier $T_z\!=\!+2$ nuclei are not stable and do not allow charge exchange reaction measurements and also because the $T_{z}\!=\!-2$ nucleus \isotope[68]{Kr} decays prominently by $\beta$ delayed protons.

\begin{figure*}
		\includegraphics[width=13.2cm]{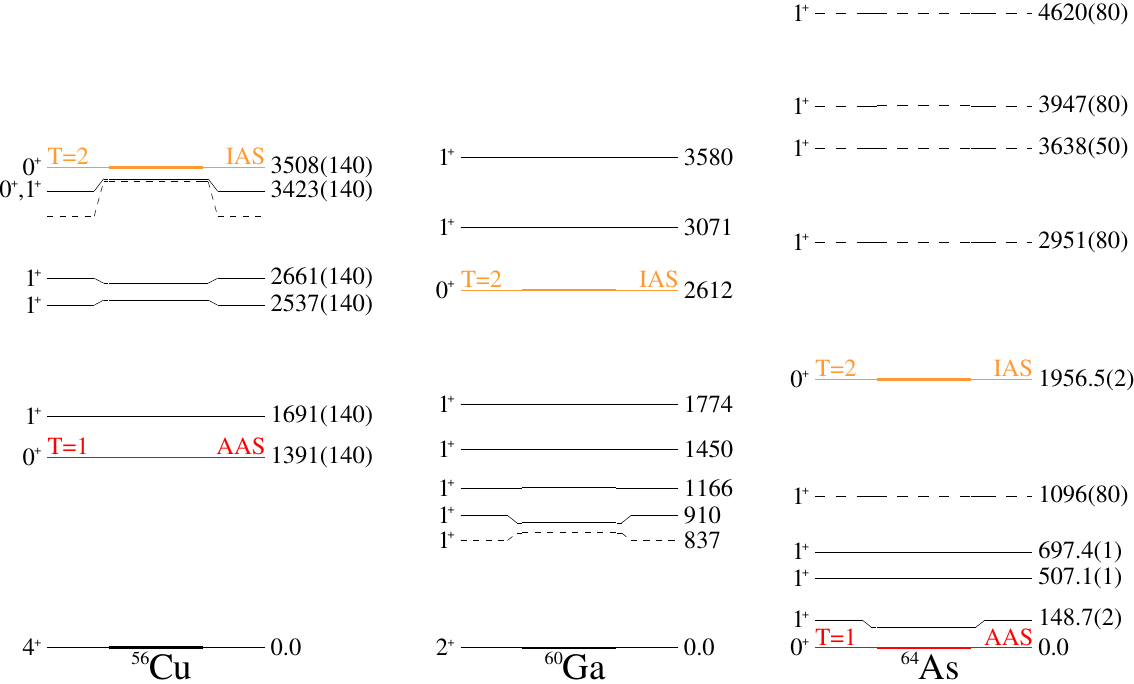}
		\caption{
			Comparison of nuclear levels in the $T_z\!=\!-1$ odd-odd nuclei populated in the $\beta$ decay of the
			corresponding $T_z\!=\!-2$ parents. The clearly identified $T\!=\!2$ IASs are marked in orange. The proposed $T\!=\!1$
			AASs are marked in red in \isotope[56]{Cu} and \isotope[64]{As}. They are not directly populated in the $\beta$ decay. See text. Levels marked by dashed 
			lines in \isotope[64]{As} indicate decay by protons and therefore their energy is not so well determined.
			\label{fig:AAS_justification}}
	\end{figure*}

\begin{figure}[!t]
		\includegraphics[width=8.6cm]{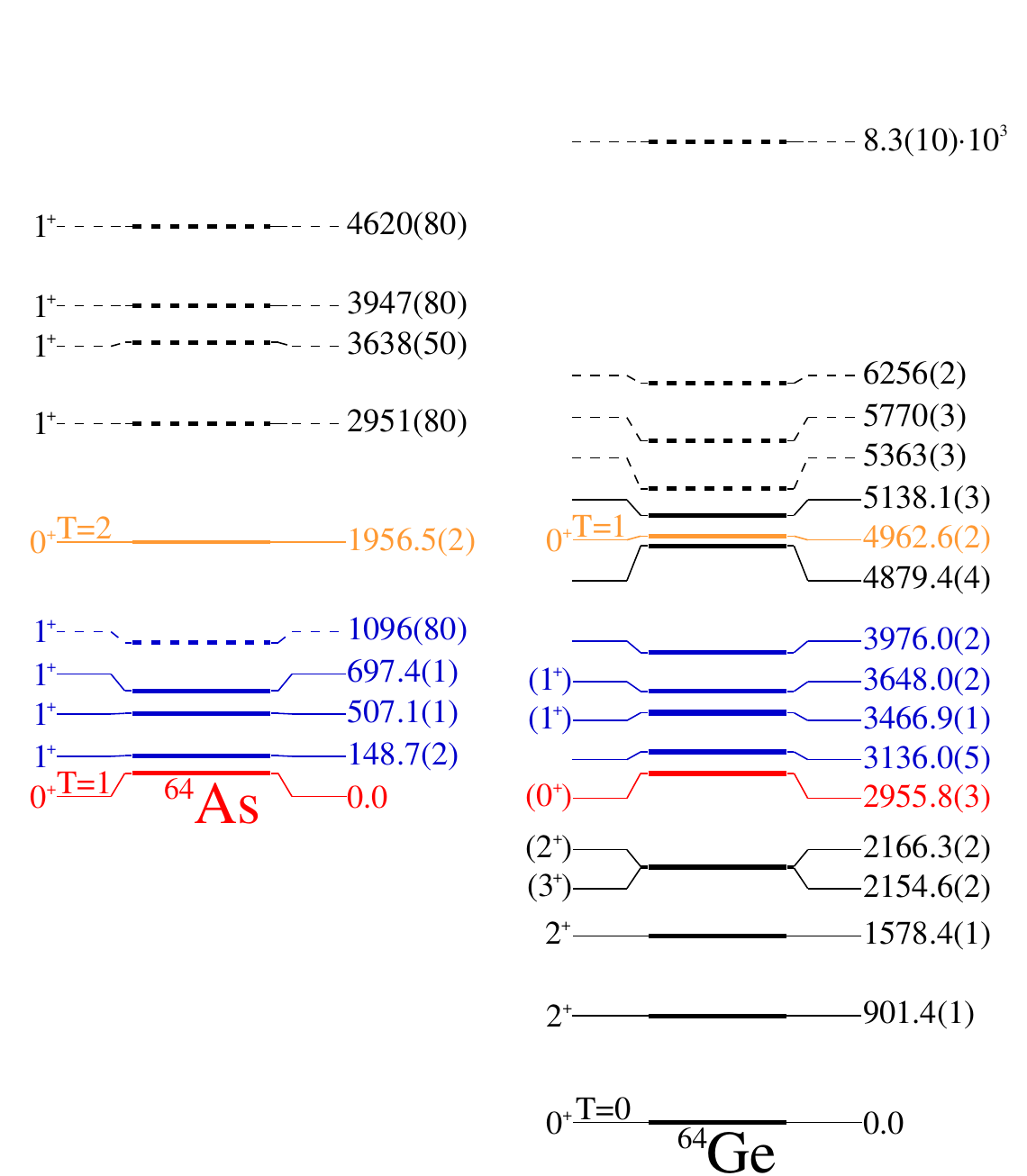}
		\caption{Comparison of the states populated in the $\beta$ decay of \isotope[64]{Se} into \isotope[64]{As} and the decay of \isotope[64]{As} into \isotope[64]{Ge}. The two IASs, $T\!=\!2$ in \isotope[64]{As} and $T\!=\!1$ \isotope[64]{Ge} have been aligned in order to show the similarities, see text.\label{fig:AAS_decay}	}
	\end{figure}

	As discussed earlier the g.s. of \isotope[64]{As} is a $0^+$ $T\!=\!1$ state. We think this is the Anti Analogue State (AAS) to the IAS. By this we mean that it is a state with very similar structure to the IAS but with one unit less in isospin. A state similar to this was studied in \isotope[56]{Co} in the 1970s using transfer reactions. It lies at $1451\mathrm{~keV}$ excitation energy and it was observed  in a two step process using the $(\isotope[3]{He},t)$ reaction at low energy \cite{rickertsen_two-step_1975}.  In the study of \isotope[56]{Zn} decay into \isotope[56]{Cu}
	\cite{Orr14}, the mirror nucleus of \isotope[56]{Co}, a similar level was observed  at
	$1391(140)\mathrm{~keV}$ excitation energy. It was indirectly populated in the de-excitation of the IAS
	\cite{Orrigo2014}.
	In Fig. \ref{fig:AAS_justification} 
	we show the trend of levels populated in the decay of the $T_{z}\!=\!-2$  nuclei \isotope[56]{Zn}, \isotope[60]{Ge} and \isotope[64]{Se}. Looking at the IAS states it is clear that this level drops in energy as the atomic number increases. We also see that the AAS $0^+$ state in \isotope[56]{Cu} drops about the same amount of energy in \isotope[64]{As} becoming the g.s. in this case.
	If this assumption is correct, then one would expect the decay of the AAS $0^+$ g.s. in \isotope[64]{As}
	to present a certain similarity with the decay of the \isotope[64]{Se} $0^+$ g.s. This is apparently the case as shown in Fig. \ref{fig:AAS_decay}, where we have aligned the $T\!=\!2$ and the $T\!=\!1$ IASs in \isotope[64]{As} and \isotope[64]{Ge}, respectively. The states lying below them show a very similar pattern, including a good candidate for the $T\!=\!0$ AAS of the $T\!=\!1$ IAS
	state in \isotope[64]{Ge} at 2955.8 keV. 
	
	%%%%%%%%% fig 22   antianalog justification %%%%%%%%%%%%%%%%%%%%
	
	%%%%%%%%% fig 23   64Se antianalog decay  %%%%%%%%%%%%%%%%%%%%

	%****************************************************
	\section{ The decay of {\isotope[\textbf{63}]{\textbf{G\lowercase{e}}}} into its mirror
		\isotope[\textbf{63}]{\textbf{G\lowercase{a}}}}
	%****************************************************
	\label{sec:analysis_63ge}
	%****************************************************
	
	Populated in the beta-delayed proton emission of \isotope[64]{Se}, the $ T_z=-1/2 $ nucleus \isotope[63]{Ge} was also identified and implanted in the WAS3ABi active stopper during the experiment, allowing us to study its beta decay. 
	The resulting  decay scheme is shown in Fig. \ref{fig:ge63_levsch}.
	The beta decay of \isotope[63]{Ge} with $ T_z\!=\!-1/2 $ populates states in its $ T_z\!=\!+1/2 $  mirror nucleus \isotope[63]{Ga}.
	The ground state of \isotope[63]{Ge} was previously assigned spin and parity $ (3/2^-) $ based on isospin symmetry \cite{ENSDF63GE}. The strong beta feeding to the
	$ 3/2^- $ ground state of \isotope[63]{Ga} observed in our experiment, makes the spin-parity of \isotope[63]{Ge} ground state $ 3/2^- $ firm (see below).
	In this case, the Fermi transition  populates the IAS with $ J^\pi\!=\!3/2^- $ and $ T\!=\!1/2 $ ground state while Gamow-Teller transitions may populate states with $ J^\pi=1/2^-,3/2^-,5/2^- $, meaning that the feeding of the IAS can also have a GT contribution. We identify the ground state of \isotope[63]{Ga} as the \isotope[63]{Ge} IAS because of the strong beta feeding to this state (see Fig. \ref{fig:ge63_levsch}). This is also what one would expect from the mirror symmetry point of view. We note here that Fermi transitions between mirror nuclei, normally called superallowed F transitions in this context, have been used to determine the $V_{ud}$ matrix element of the CKM matrix \cite{severijns_f_2023}. We present here two main contributions to this discussion. Firstly we obtained a more precise value of the $T_{1/2}$ of the \isotope[63]{Ge} ground state. 
	The previous values were $150(9)\mathrm{~ms}$ \cite{blank2002}, $149(4)\mathrm{~ms}$ \cite{rogers2014} and $156(11)\mathrm{~ms}$ \cite{kucuk2017} while we obtained $152.6(3)\mathrm{~ms}$. Secondly, we obtained the first experimental determination of the ground state to ground state  transition, which has $90.1(3)\%$ intensity. Assuming $B(F)\!=\!1$, we obtain a value of $81.3(3)\%$ for the Fermi contribution. Similar values have been obtained for the decays of other mirror nuclei of similar mass (see Table II in \cite{severijns_f_2023}).
	Excited states in \isotope[63]{Ga} have only been studied in in-beam experiments \cite{Balamuth1991,Weiszflog2001}. Three of these states were also observed in this experiment and are shown in black in Fig. \ref{fig:ge63_levsch}. 
	Based on $\gamma$-$\gamma$ coincidences we found 
	seven new excited states in \isotope[63]{Ga}. The known levels in \isotope[63]{Ga} are the $75.0 \mathrm{~keV}$ $5/2^-$, the $442.8 \mathrm{~keV}$ $3/2^-$ 
	and $722.2 \mathrm{~keV}$ assigned ($5/2^-$) in \cite{rudolph_2021}. In the same paper, several shell model calculations show that a $1/2^-$ state is expected at low excitation energy in \isotope[63]{Ga}. An excellent candidate for this state is the newly observed level at $627.0 \mathrm{~keV}$, which decays only to the $3/2^-$ ground state and not to the first excited $5/2^-$ state only $75.0 \mathrm{~keV}$ above. We note here that a $\gamma$-ray with similar energy, namely $624.7 \mathrm{~keV}$ has been observed in the previous in-beam studies \cite{Balamuth1991,Weiszflog2001,rudolph_2021}, but we believe it is not the same transition, not only because of the slightly different energy but also because of the location in the level scheme de-exciting a $9/2^-$ level \cite{rudolph_2021}. 
	
	In summary this work has also contributed to a better knowledge and understanding of the beta decay between the $T_z\!=\!-1/2$ \isotope[63]{Ge} and  the $T_z=+1/2$ \isotope[63]{Ga} mirror nuclei.

	%\clearpage
	%****************************************************
	\section{Conclusions}
	%****************************************************
	\label{sec:conclusions}
	%****************************************************
	
	In the present work we have reported on our studies of the decay of three exotic nuclei, namely \isotope[64]{Se}, \isotope[64]{As} and \isotope[63]{Ge}. In two of them, \isotope[64]{Se} and \isotope[64]{As}, we observed beta-delayed proton emission. 
	The experiment allowed us to measure for the first time the half-life of \isotope[64]{Se}, and measure  half-life values for the other isotopes with slightly better precision than previously reported. 
	We also observed the proton spectrum for \isotope[64]{Se} and \isotope[64]{As} decays and determined the corresponding proton branching ratios, $48.0(9)\%$ and $ 4.4 (1) \% $ respectively.
	
	Using the $\beta$ delayed proton radiation and the $\beta$ delayed $\gamma$ radiation after \isotope[64]{Se} decay, we could extract an experimental value for the ground state mass of \isotope[64]{As}. The value deduced agrees with a recent direct mass measurement performed at Lanzhou with 6 event statistics and has a much reduced error. We could also deduce the ground state mass of \isotope[64]{Se} using the IMME for the $ T\!=\!2 $ multiplet.
	
	Detailed decay schemes could be constructed thanks to the high intensity of the \isotope[78]{Kr} primary beam produced at the RIKEN Nishina Centre.  Combining these results with charge-exchange reaction data from RCNP allowed a very detailed description of the A=64 $ T\!=\!2 $ multiplet including the first observation of the IAS in \isotope[64]{As}. This is the heaviest 
	$ T\!=\!2 $ multiplet where these studies are possible. A good mirror symmetry was observed between the  $ T_z\!=\!1 $ and $ T_z\!=\!-1 $ $ A\!=\!64 $  nuclei. 
	
	New experimental information  was obtained for excited states in the  N=Z nucleus \isotope[64]{Ge}, previously known only from in-beam studies.
	The quality of the data allowed one to revisit the idea of the AAS (Anti Analogue State), which helped the interpretation of these decays.
	
	Moreover, thanks to the beta delayed proton decay, an excited state was observed at $417.5\mathrm{~keV}$ in the exotic \isotope[63]{Ge} nucleus which is in good agreement with its mirror state in \isotope[63]{Ga}.
	It should be noted that the \isotope[63]{Ge} nucleus, cannot be reached in the $\beta$ decay of \isotope[63]{As}, because this nucleus is probably unbound. 
	
	The beta decay of \isotope[63]{Ge} into its mirror nucleus \isotope[63]{Ga} was studied for the first time allowing for a firm determination of its ground state spin-parity. The beta feeding of the superallowed F plus GT transition to the IAS could be obtained together with a more precise value of the ground state $T_{1/2}$.
	
	In summary, in this study, a large amount of new and interesting spectroscopic results has been presented which brings detailed decay studies very close to the proton drip line.

	%\clearpage
	%****************************************************
	%**** Acknowledgements
	%****************************************************
	
	\begin{acknowledgments}
		This experiment was performed at RI Beam Factory operated by RIKEN Nishina Center and CNS, University of Tokyo.
		We want to express our gratitude to the RIBF accelerator staff for providing a stable and high-intensity \isotope[78]{Kr} beam during the experiment, the EUROBALL Owners Committee for the loan of germanium detectors, and the PRESPEC Collaboration for the readout electronics of the cluster detectors. 
		Part of  WAS3ABi was supported by the Rare Isotope Science Project, which is funded by the Ministry of Science, ICT and Future Planning (MSIP) and National Research Foundation (NRF) of Korea.
		Support from the Spanish Ministry  under Grants No. IJCI-2014-19172, No. FPA2011-24553, No. FPA2014-52823-C2-1-P and No. FPA2017-83946-C2-1-P, PID2022-138297NB-C21 
		Centro de Excelencia Severo Ochoa del IFIC SEV-2014-0398, 
		and Junta para la Ampliación de Estudios Programme (CSIC JAE-Doc contract) co-financed by FSE is acknowledged. 
		We acknowledge the support of the Generalitat Valenciana Grant No. PROMETEO/2019/007 and CIPROM/2022/9. 
		This work was also financed by Programmi di Ricerca Scientifica di Rilevante Interesse Nazionale (PRIN) No. 2001024324 01302, 
		by JSPS KAKENHI Grant No. 25247045, 2604808, 
		by the Chilean Programme of Postgraduate Fellowship CONICYT (Grant D-21151450), 
		by the programme i+COOP+2015 N. COOPA20125 (Spain), 
		by the project FONDECYT REGULAR N. 1171467, 1221364, ANID - Millennium Science Initiative Program - ICN2019\_044, 
		by the Istanbul University Scientific Project Unit under Project No. BYP-53195, 
		by the NSF Grant No. PHY-1404442, 
		by the UK Science and Technology Facilities Council (STFC) Grant No. ST/F012012/1, 
		by MEXT Japan under Grant No. 15K05104, 
		by the French-Japanese Laboratoire international associ FJ-NSP, 
		and by NKFIH (NN128072) and NKFIH (K1470109), by the New National Excellence Program of the Ministry for Innovation and Technology (ÚNKP-19-4-DE-65). 
		R.B.C. acknoledges the support by the Max-Planck Partner group, 
		P.-A. Söderström acknowledge the contract PN 23.21.01.06 sponsored by the Romanian Ministry of Research, Innovation and Digitalization.
	\end{acknowledgments}

	%****************************************************
	%**** Biblio
	\bibliography{references.bib}

\end{document}
%
% ****** End of file apstemplate.tex ******